%% file: main.tex
\documentclass[sttt]{svjour}

\usepackage{amsmath}
\usepackage{amssymb}
\usepackage{array}
\usepackage{temporal}
\usepackage{enumerate}
\usepackage{graphicx}
\usepackage{bbm}
\usepackage{cite}
\usepackage{tikz}
\usepackage{paralist}
\usepackage{fixme,manfnt,mparhack}
\usepackage[T1]{fontenc}
\usepackage[ruled,linesnumbered]{algorithm2e}
\usepackage{xspace}
\usepackage{hyperref}
\usepackage{wrapfig}

\usepackage{macros}

\usepackage{makeidx}  
\usepackage{color}
\usepackage{amsfonts}
\PassOptionsToPackage{hyphens}{url}
\usepackage{listings}
\usepackage{subfigure}

\usepackage{booktabs}

\usepackage{rotating}
\usepackage{algpseudocode}

\input{preamble.tex}

\title{The First Reactive Synthesis Competition\\ (\syntcomp 2014)}
\author{Swen Jacobs\inst{1,2}  \and Roderick Bloem\inst{1} 
				\and Romain Brenguier\inst{3} 
				\and R\"udiger Ehlers\inst{4,5}
                                \and Timotheus Hell\inst{1}
				\and Robert K\"onighofer\inst{1} 
				\and Guillermo A. P\'erez\inst{3} 
				\and Jean-Fran\c{c}ois Raskin\inst{3} 
				\and Leonid Ryzhyk\inst{6,7} 
				\and Ocan Sankur\inst{8,3} 
				\and Martina Seidl\inst{9} 
				\and Leander Tentrup\inst{2} 
				\and Adam Walker\inst{6} }
\institute{Graz University of Technology, Austria 
						\and Saarland University, Saarbr\"ucken, Germany 
						\and Universit\'e Libre de Bruxelles
						\and University of Bremen, Germany 
						\and DFKI GmbH, Bremen, Germany 
						\and NICTA, Sydney, Australia 
						\and Carnegie Mellon University,
                                                Pittsburgh, USA
						\and CNRS, IRISA, Rennes, France
						\and Johannes-Kepler-University Linz, Austria}
\begin{document}
\maketitle

\begin{abstract}
We introduce the reactive synthesis competition (\syntcomp), a long-term effort intended to stimulate and
guide advances in the design and application of synthesis procedures for reactive
systems. 
%
The first iteration of \syntcomp is based on the controller synthesis problem for finite-state systems and safety 
specifications. We provide an overview of this problem and 
existing approaches to solve it, and report on the design and results of the 
first \syntcomp. This includes the definition of the benchmark format, the 
collection of benchmarks,
the rules of the competition, and the five synthesis tools that participated.
We present and analyze the results of the competition and draw conclusions on the state of the art.
Finally, we give an outlook on future directions of \syntcomp.
\end{abstract}

\input{intro}

\input{problem}

\input{format}
\input{benchmarks}

\input{rules}

\input{tools}

\input{execution}

\input{results}


\input{conclusions}

\bibliographystyle{abbrv}
\bibliography{synthesis}
\end{document}

%% file: preamble.tex
\newcommand{\mysubsubsection}[1]{\medskip \noindent {\bf #1.}}
\newcommand{\myparagraph}[1]{\smallskip \noindent {\em #1.}}

\newcommand{\bbB}{\mathbb{B}}

\newcommand{\depqbf}{\textsf{DepQBF}\xspace}
\newcommand{\bloqqer}{\textsf{Bloqqer}\xspace}
\newcommand{\demiurge}{\textsf{Demiurge}\xspace}
\newcommand{\learn}{\textsf{(learn)}\xspace}
\newcommand{\templ}{\textsf{(templ)}\xspace}
\newcommand{\dparallel}{\textsf{(parallel)}\xspace}
\newcommand{\ltltoaig}{\texttt{ltl2aig}\xspace}
\newcommand{\syntcomp}{SYNTCOMP\xspace}
\newcommand{\minisat}{\textsf{Minisat}\xspace}
\newcommand{\lingeling}{\textsf{Lingeling}\xspace}
\newcommand{\basil}{\textsc{Basil}\xspace}
\newcommand{\abssynthe}{AbsSynthe\xspace}
\newcommand{\simpleBDD}{Simple BDD Solver\xspace}
\newcommand{\realizer}{Realizer\xspace}

\newcommand{\Abc}{\textsf{ABC}\xspace}

\newcommand{\subby}{\kern-0.2em\leftarrow\kern-0.2em}

\newcommand{\secref}[1]{Section~\ref{#1}}

%% file: intro.tex
\section{Introduction}

Ever since its definition by Church~\cite{Church62}, the automatic synthesis 
of reactive systems from formal specifications has been
one of the major challenges of computer science, and an active field 
of research. A number of fundamental approaches to solve the problem have 
been proposed (see
e.g.~\cite{EmersonC82,Rabin69,PnueliR89}). Despite the obvious advantages of
automatic synthesis over manual implementation and the significant progress of
research on theoretical aspects of synthesis, the impact of formal
synthesis procedures in practice 
has been very limited. One reason for this limited impact is the scalability
problem that is inherent to synthesis approaches. The reactive synthesis
problem is in general 2EXPTIME-complete for LTL specifications~\cite{PnueliR89}.
A number of approaches have recently been invented to solve special cases of 
the problem more efficiently, either by restricting the specification
language~\cite{BloemJPPS12}, or by a smart exploration of the search
space~\cite{Finkbeiner13,Ehlers12,fjr11,SohailS13,FinkbeinerJ12,FiliotJR13}. 
While important progress on
the scalability problem has been made, an additional problem is the lacking
maturity and comparability of implementations, and a lack of incentive for 
the development of efficient implementations~\cite{Ehlers11a}. 
Solving different aspects of 
this problem is the main motivation of \syntcomp, as explained in the following
(inspired by~\cite{LecoutreRD10}).

\myparagraph{Synthesis tools are hard to compare}
Research papers that introduce a new algorithm in many cases do include a
comparison of its implementation against existing ones. However, the 
comparison of a large number 
of tools on a benchmark set of significant size can take weeks or months of 
computation time.
This is often circumvented in research papers by comparing the new
results to existing experimental data (usually obtained under different
experimental conditions), or by comparing against a small number of tools
on a small benchmark set. In both cases, this limits the value of the
experimental results. In contrast, \syntcomp provides reliable results for a 
significant number of synthesis tools on a large
benchmark set, with consistent experimental conditions. 

\myparagraph{It is hard to exchange benchmark sets}
Related to the comparison of tools, we note that almost every existing tool uses
its own input language, and benchmarks have to be translated from one format to
another in order to compare different tools. This makes it hard to exchange
benchmark sets, and adds another source of uncertainty when comparing tools. 
\syntcomp aims to solve these issues by defining a standard benchmark format,
and by collecting a benchmark library that is publicly available for the
  research community. 

\myparagraph{Usability of synthesis tools}
Implementations of many synthesis approaches do
exist~\cite{Ehlers11,bbfjr12,BloemCGHKRSS10}, but they
cannot effectively be used as \emph{black-box solvers for
  applications}. The definition of a standard language is a first step in this
direction. In addition, the competition forces tool developers to produce
implementations that are sufficiently robust to work on the complete benchmark
library of \syntcomp with a fixed configuration. Thus, \syntcomp promotes the
\emph{simplicity of use} that comes with push-button
approaches that do not require any user intervention.

\medskip
\noindent
Summing up, the goal of the \emph{reactive synthesis competition} (\syntcomp) is
to foster research in scalable and user-friendly implementations of synthesis
techniques. 

\mysubsubsection{Related competitions} 
Competitions have been used to achieve these goals in many related
fields, including automated reasoning~\cite{SutcliffeS06,BarrettMS05,JarvisaloBRS12}
and automated verification~\cite{Beyer12}\footnote{See also: the Hardware Model Checking Competition, \url{http://fmv.jku.at/hwmcc/.} Accessed February 2016.}. 
A difference of synthesis competitions to most of the 
competitions in automated reasoning or verification is that solutions to the 
synthesis problem can be ranked according to inherent quality criterions that go beyond mere correctness, 
such as reaction time or size of the solution. Thus, a synthesis competition also needs 
to measure the quality of solutions with respect to these additional metrics. 

In parallel to \syntcomp 2014, the \emph{syntax-guided synthesis competition} 
(SyGuS-COMP) was held for the first time~\cite{AlurBDF0JKMMRSSSSTU15}. The focus of 
SyGuS-COMP is on the synthesis of functional instead of reactive programs, 
and the specification is given as a first-order logic constraint on the 
function to be synthesized, along with a syntactic constraint that restricts 
how solutions can be built. The goals of SyGuS-COMP are similar to those of 
\syntcomp, but for a fundamentally different class of programs and specifications.

\mysubsubsection{Timeline} 
The organization of the first \syntcomp began formally with a presentation and 
discussion of ideas at the second Workshop on Synthesis (SYNT) in July 2013.  
The organization team consisted of Roderick Bloem, R\"udiger Ehlers and Swen 
Jacobs. The decision for the specification format was made and announced in 
August 2013, and a call for benchmarks, along with the rules of the 
competition, was published in November 2013. In March 2014 we published our 
reference implementation, and benchmarks were collected until the end of 
April 2014. Participants had to submit their tools until the end of May 2014, 
and the experiments for the competition were executed in June and July 2014. 
The results were first presented at the 26th International Conference on 
Computer Aided Verification (CAV) and the 3rd SYNT Workshop in 
July 2014. 

\mysubsubsection{Goals} 
The first competition had the following goals: 
\begin{itemize} 
\item define a class of synthesis problems and a benchmark format that results in a low 
entry-barrier for interested researchers to enter the competition 
\item collect benchmarks in the \syntcomp format 
\item encourage development of synthesis tools that support the \syntcomp format 
\item provide a lobby that connects tool developers with each other, and with   possible users of synthesis 
tools \end{itemize}
\syntcomp 2014 was already a success before the experimental evaluation began:
within less than $10$ months after the definition of the benchmark format,
we collected $569$ benchmark instances in $6$ classes of benchmarks, and $5$
synthesis tools from $5$ different research groups were entered into the competition. 
For four of the tools, at least one of the
developers was present at CAV and/or the SYNT workshop.

\mysubsubsection{Overview} 
The rest of this article describes the background, design, participating solvers (called \emph{entrants}), and
results of \syntcomp 2014. We will introduce the synthesis problem for safety
specifications, as well as different approaches for solving it, in
Section~\ref{sec:problem}. We define the \syntcomp format in
Section~\ref{sec:format}, and describe the benchmark set for \syntcomp 2014 in
Section~\ref{sec:benchmarks}. Section~\ref{sec:rules} defines the rules of the
competition. In Section~\ref{sec:participants} we give an overview of the
entrants of \syntcomp 2014, followed by some notes on the execution of the
competition in Section~\ref{sec:execution}. In Section~\ref{sec:results}, we
present and analyze the experimental results of the competition.

Note that Sections~\ref{sec:benchmarks} and~\ref{sec:participants}, as well as
parts of Section~\ref{sec:problem} are based on the descriptions that the
respective benchmark and tool authors had to supply in order to participate. 
The setup of the competition framework as described in
Section~\ref{sec:execution} was taken care of by Timotheus Hell. The
remainder of this article is original work of the \syntcomp organizers.

This article is based on the first description of the \syntcomp format~\cite{SYNTCOMP-format} and a preliminary version of the \syntcomp 2014 report~\cite{Jacobs15}.

%% file: problem.tex
\section{Problem Description and Synthesis Approaches}
\label{sec:problem}

Informally, the reactive synthesis problem consists of finding a system $C$ 
that satisfies a given specification $\varphi$ 
in an adversarial environment. In general, systems may be infinite-state 
(programs) or finite-state (circuits), and specifications can come in 
different forms, for example as temporal logic formulas or as monitor circuits.

For the first \syntcomp, we aimed for a low entry-barrier for 
participants, and to keep the competition manageable in terms of tasks like 
the definition of input and output format, and the verification of results. 
To this end, we only consider the 
synthesis of finite-state systems from pure safety specifications modeled as 
monitor circuits. The monitor circuit reads two kinds of input signals: 
uncontrollable inputs from the environment, and controllable inputs from the 
system to be synthesized. It raises a special output $\out$ if the safety 
property $\varphi$ has been violated by the sequence of input signal 
valuations it has read thus far.

Then, the \emph{realizability problem} is to determine if there exists a
circuit $C$ that reads valuations of the uncontrollable inputs and provides
valuations of the controllable 
inputs such that $\out$ is not raised on any possible execution. The
\emph{synthesis problem} is to provide such a $C$ if it exists.
As a quality criterion, we consider the \emph{size} of the produced
  implementation, which not only correlates to the cost of implementing a
  circuit in hardware, but often also leads to implementations which have other
  desirable properties, like short reaction time.

\subsection{Synthesis as a Safety Game}

The traditional approach to reactive synthesis is to view the problem 
as a game between two players~\cite{BL69,Rabin69,Thomas95}: the \emph{environment player}
decides uncontrollable inputs, and the \emph{system player} decides controllable
inputs of the monitor circuit. \emph{States} of the game are valuations of the
latches in the monitor circuit. A
state is \emph{safe} if $\out$ is not raised. The goal
of the system player is to visit only safe states, regardless of the
environment behavior.

\mysubsubsection{Game-based Synthesis}
In a first step, a so-called \emph{winning region} for the system player is 
computed. The winning region $W$ is the set of all states from which the system 
player can enforce the specification, i.e., from which it can guarantee that the
environment cannot force the game into an unsafe state.

In a second 
step, a \emph{winning strategy} is derived from the winning region.  For 
every state and every valuation of the uncontrollable inputs, the winning strategy
defines a set of possible valuations of the controllable inputs that can ensure
that the winning region is not left.

The last step is to implement this strategy in a circuit, where a concrete
choice for the controllable inputs has to be made for every state and valuation
of uncontrollable inputs.

All of the tools in \syntcomp 2014 implement such a game-based synthesis 
approach, in one form or another.

\mysubsubsection{Symbolic encoding}
To achieve acceptable scalability, it is important to implement synthesis 
algorithms symbolically, i.e., by manipulating formulas instead of enumerating 
states~\cite{AlurMN05}. In synthesis, symbolic algorithms are usually
implemented with Binary  Decision Diagrams
(BDDs)~\cite{Bryant86,somenzi99}. Most of the tools in \syntcomp 2014 use
BDD-based approaches with different optimizations to achieve good performance in 
synthesis time and circuit size.

However, BDDs also have scalability issues, in particular the growing size of
the data structure itself. Alternatively, the problem
can be encoded into a sequence of propositional satisfiability (SAT), quantified 
Boolean formulas (QBF), or satisfiability modulo theories (SMT) problems. The 
enormous performance improvements in decision procedures for satisfiability 
over the last decades encourage such approaches. 

In the following, we give a mostly informal description of the 
three synthesis techniques used by the tools that entered 
\syntcomp 2014: BDD-based game solving (\secref{sec:BDDgame}), 
SAT-/QBF-based game solving (\secref{sec:SATgame}), and 
template-based synthesis (\secref{sec:templateApproach}).

\subsection{Preliminaries: Circuits and Games}
Let $\mathbb{B} = \{0,1\}$. If~$X$ denotes a finite set of Boolean variables,
then any $v \in \bbB^X$ is called a \emph{valuation of $X$}. 
Sets of valuations of $X$ are represented by 
\emph{quantified Boolean formulas} on $X$, which are made of propositional logic and
first-order quantification on~$X$.  A~\emph{formula}~$f$ with free variables
$X$ will be written as~$f(X)$, and for the same formula under a given valuation $v$
of $X$ we write $f(v)$.  If the free variables are~$X\cup Y$, 
we also write $f(X,Y)$.  For a set of variables~$X=\{x_1,\ldots,x_n\}$, we 
write $\exists X$ instead of $\exists x_1\exists x_2 \ldots \exists x_n$, and 
similarly for universal quantification. For a set of variables $X =
\{x_1,\ldots,x_n\}$, we use $X'$ to denote $\{x_1',\ldots,x_n'\}$, a set of
primed copies of the variables in $X$, usually representing the variables
after a step of the transition relation.

Then, the synthesis problem is given as a (sequential) monitor circuit $M$ over
the sets of variables $L$, $X_u$, $X_c$, where
\begin{itemize}
\item $L$ are state variables for the latches in the monitor circuit,
\item $X_u$ are uncontrollable input variables,
\item $X_c$ are controllable input variables, and
\item $\out \in L$ is a special variable for the unsafe states, i.e., a state is
  unsafe iff $\out =1$.\footnote{For simplicity, we assume that $\out$ is a
    latch. If $\out$ is not a latch in the given monitor circuit, then it can
    be described as a formula $f(L,X_u,X_c)$. In this case we can obtain a
    problem in the described form by 
    extending the circuit with a new latch that takes $f(L,X_u,X_c)$ as an input
    and provides output $\out$.}
\end{itemize}
We assume that the system has a unique initial state, in which all latches in
$L$ including $\out$ are initialized to $0$. 

\begin{figure}
\input{problem_fig}
\caption{Synthesis problem with monitor circuit $M$ and (unknown) system circuit
  $C$}
\label{fig:problem}
\end{figure}
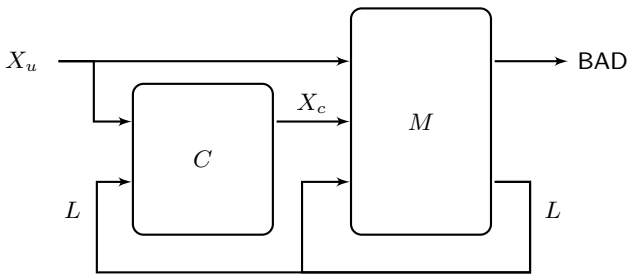

A \emph{solution} of the synthesis problem is a sequential circuit $C$ with
inputs $L \cup X_u$ and outputs $X_c$, such that the composition of $C$ and $M$
is safe, i.e., states with $\out = 1$ are never reached, for any sequence of
(uncontrollable) inputs $X_u$ and starting from the unique initial state. The
synthesis problem is depicted in Figure~\ref{fig:problem}.

Note that a circuit defines a (Mealy-type) \emph{finite-state machine} in the 
standard way. With the additional distinction between controllable and 
uncontrollable inputs and the interpretation of $\out$ as the set of unsafe 
states, it defines a \emph{safety game}: the set of states is $\bbB^{L}$ (
valuations of latches $L$), with initial state $0^{L}$. The transition 
relation of the monitor circuit can be translated into a formula $T(L, X_u, X_
c, L')$ that relates valuations of $L,X_u,X_c$ to the valuation of (next-
state variables) $L'$. In every turn of the game, first the environment 
player chooses a valuation of $X_u$, and then the system player chooses a 
valuation of $X_c$. The successor state is computed according to $T(L, X_u, X_
c, L')$. A \emph{strategy} of the system player is a function that maps the 
sequence of valuations of $L$ and $X_u$ seen thus far to a set of possible 
valuations for $X_c$. It is \emph{deterministic} if it always maps to a 
unique valuation. A strategy is \emph{winning} for the system player if it 
avoids entering the unsafe states regardless of the actions of the 
environment. Two-player safety games are \emph{determined}, i.e., for every 
such game either the environment player or the system player has a winning 
strategy. A \emph{memoryless} strategy only depends on the current values of $
L$ and $X_u$. For safety games, there exists a winning strategy iff there 
exists a memoryless winning strategy. A deterministic memoryless winning 
strategy can be represented as a circuit, and thus provides a solution to the 
synthesis problem.

\subsection{BDD-based Game Solving}
\label{sec:BDDgame}


For a basic BDD-based algorithm, assume that the transition relation
$T(L, X_u, X_c, L')$ and the sets of initial and unsafe states are each
represented as a single BDD (see e.g.~\cite{cm90}).
To determine whether the environment has a strategy that guarantees
its victory, one repeatedly computes the set of states from which it can force
the game into the unsafe states. 
If $S(L')$ is a formula over the latches $L'$, representing a set of states,
then the set of \emph{uncontrollable predecessors} of $S(L')$ can be computed as
the set of valuations of latches $L$ that satisfy
$$\upre(S(L'))= \exists X_u\ \forall X_c\ \exists L'.\ S(L') \wedge T(L,X_u, X_c, L').$$
To compute the winning region $W(L)$ of the system player, we first compute the
least fixpoint of $\upre$ on $\out$:
\begin{equation} \label{eq:fixpoint}
\mu S(L). \; \upre(S(L') \lor \out')
\end{equation}
The resulting set of latch valuations represents the states from which
the environment can force the game into the unsafe states. Since two-player
safety games are determined, the complement of this set is the winning region
$W(L)$ for the system player (see, e.g.~\cite{Thomas95}).

That is, if during our fixpoint computation we notice that the environment can
force the game from the initial state to the unsafe states, then we can stop ---
the specification is unrealizable. Otherwise, the initial state will be
contained in the winning region $W(L)$, and $W(L)$ represents
a non-deterministic strategy 
$\lambda$ for the system player, which can be described as a function $\lambda$ that maps a valuation $s \in \bbB^L$ of the latches and a valuation $\sigma_u \in \bbB^{X_u}$ of the uncontrollable variables to a set of possible valuations for the controllable variables: 
$$ \lambda(s,\sigma_u) = \{ \sigma_c \in \bbB^{X_c}
\mid \forall L'.\ T(s,\sigma_u,\sigma_c,L') \rightarrow
W(L')\}.
$$
To solve the synthesis problem, in principle any determinization of $\lambda$
can be chosen to obtain a functional strategy for the system player.

In order to compute the winning region efficiently and to find a strategy that
can be represented as a small circuit, a number of optimizations can be used. We introduce some of the common optimizations in the following, in order to be able to compare and distinguish the participants that use a BDD-based approach.

\mysubsubsection{Partitioned Transition Relation and Direct Substitution}
To be efficient,
the explicit construction of the BDD for the transition relation should be
avoided. This can be achieved by \emph{partitioning} 
the transition relation into a set of simpler relations, inspired by similar
approaches for the model checking problem~\cite{cm90,BurchCL91,Ranjan95}. A common
approach is to split $T(L,X_u, X_c, L')$ into a set of (functional) relations
$\{f_l(L,X_u,X_c)\}_{l \in L}$, where each $f_l$ represents the next-state value
of latch $l$.

Then, the uncontrollable predecessor can be computed as 
$$\upre(S(L')) = \exists X_u\ \forall X_c.\ S(L')[l' \leftarrow f_l(L,X_u,X_c)]_{l \in L},$$
avoiding to ever build the monolithic transition relation, as well as having 
to ever declare a next state copy of any latch variable in the BDD manager. 
Substituting individual latches with functional BDDs is directly supported by 
existing BDD packages, e.g., function \textsc{bddVectorCompose} in the CUDD 
package~\cite{somenzi99}. This will be called \emph{partitioned transition
  relation} in the tool descriptions.

A special case of this approach is to identify those latches that only store the value 
of some other variable from the last step, i.e., the latch update function 
has the form $f_l = x$ for some $x \in L \cup X_u \cup X_c$, and use the
substitution of latches by $f_l$ only for them. In this case, we only substitute
with a single existing variable instead of a functional BDD, which can be done,
e.g., with CUDD's \textsc{bddVarMap} method. This will be called \emph{direct
  substitution} in the tool descriptions.

\mysubsubsection{BDD Reordering}
Efficiency of BDD-based algorithms depends to a large extent on the size of 
BDDs, which in turn depends on the variable ordering~\cite{AzizTB94}. To keep
the data structures small, \emph{reordering} heuristics are 
commonly used to try to minimize BDDs at runtime~\cite{Rudell93}.
Standard BDD packages come 
with pre-defined reordering strategies. Algorithms that do not use reordering 
at all are usually not competitive.

\mysubsubsection{Efficient Computation of $\upre$}
In the fixpoint computation, we repeatedly use the $\upre$ operation to
compute the set from which the environment wins the game. This operation
consists of conjoining the current set of states with the transition relation,
followed by resolving the quantification over inputs and current states to get
the description of the set of predecessor states. The latter is called
(existential or universal) \emph{abstraction} over these variables. In practice,
it is often preferable to not use this strict order, but instead do conjunction
and abstraction in parallel. 
This is directly supported in some BDD packages, e.g., in CUDD's
\textsc{bddAndAbstract} method. We will call this optimization \emph{simultaneous 
  conjunction and abstraction}.

\mysubsubsection{Eager Deallocation of BDDs}
Another optimization is to deallocate BDDs that are no longer needed as soon as
possible. Not only do these BDDs take up memory, but more importantly they are also
counted and analyzed for BDD reordering. Thus, removing such BDDs
saves space and time. We will call this \emph{eager deallocation}.

\mysubsubsection{Abstraction-based algorithms}
For systems that have a large state space,
an \emph{abstraction approach} may be more efficient 
than precise treatment of the full state space~\cite{CousotC77,GrafS97,HenzingerJM03}. 
This can be done by defining 
a set of predicates $P$ (which may simply be a subset of the state variables 
of the system~\cite{kurshan94,dealfaro}) and computing over- and 
under-approximations of the 
\upre\ function, with respect to the partitioning of the state space defined 
by the predicates in $P$. Computing fixpoints for these approximations is usually 
much cheaper than 
computing the precise fixpoint for the system. If the system player wins the
game for the over-approximation of \upre, then it also wins the original
game. If the system player loses for the under-approximation of \upre, then 
it also loses the original game. If neither is the case, then the 
abstraction is insufficient and needs to be refined by introducing 
additional predicates.

\mysubsubsection{Extraction of Small Winning Strategies}
To obtain from $\lambda$ a functional strategy that can be represented as a
small circuit, a number of  
optimizations is commonly used. To this end, let $\lambda_c$ be the restriction
of $\lambda$ to one output $c \in X_c$. We want to obtain a partitioned
strategy, represented as one function $f_c(L,X_u)$ for every $c \in X_c$: 
\begin{itemize}
\item For every $c \in X_c$, in some arbitrary order, compute the positive 
and negative co-factors of $\lambda_c$, i.e., the values 
$s,\sigma_u$ for which $\lambda_c(s,\sigma_u)$ can be $1$ or $0$, 
respectively. 
These can be used to uniquely define $f_c$, e.g., by letting $f_c(s,\sigma_u)=
 0$ for all values in the negative co-factor, and $f_c(s,\sigma_u)=1$ 
otherwise~\cite{BloemGJPPW07}. This will be called \emph{co-factor-based 
extraction of winning strategies}.

\item
After extracting the functions $f_c(L,X_u)$ for all $c \in X_c$, one can 
minimize the strategy by doing an additional \emph{forward reachability analysis}:
compute the reachable states \emph{with this strategy}, and restricting all
$f_c$ to values of $L$ that are actually reachable.
\item After translating the functions $f_c(L,X_u)$ into an AIG 
representation, a number of minimization techniques can be used to obtain 
small 
AIGs~\cite{FRAIGs,MishchenkoCB06}. The verification tool \Abc\footnote{\url{http://www.eecs.berkeley.edu/~alanmi/abc/}. Accessed February 2016.}~\cite{ABC} 
implements a number of these minimization strategies that can be used in a 
black-box fashion to obtain smaller circuits, and we will call this approach 
\emph{\Abc   minimization}. \end{itemize}

\subsection{Incremental SAT- and QBF-based Game Solving}
\label{sec:SATgame}

In contrast to the BDD-based approaches already presented, the SAT- and QBF-based
approaches start with a coarse over-approximation of the winning region,
represented as a CNF formula $W(L)$ over the state 
variables $L$. This approximation is incrementally refined, such that $W(L)$
eventually represents the winning region 
symbolically. 

More concretely, we initialize $W(L)$ to the set of all safe
states $\neg \out$.
In each iteration, the underlying solver is used to compute a state $s \models 
W(L) \wedge \upre(\neg W(L'))$ within the current candidate version $W(L)$ of 
the winning region from which the environment player can enforce to leave $W(L)$ 
in one step. Obviously, such a state cannot be part of the winning region. 
Hence, we refine $W(L)$ by removing this state.  The state $s$ can 
be represented as a cube over the state variables $L$, so removing $s$ from $W(L)$ 
amounts to adding the clause $\neg s$.  

In order to remove a larger region from 
$W(L)$, the algorithm tries to generalize the clause $\neg s$ by removing literals, 
as long as the resulting clause $\neg \tilde{s}$ still only excludes states from 
which $\neg W(L)$ can be reached by the environment in one step.  More 
specifically, literals are dropped as long as $(W(L) \wedge \neg  \tilde{s}) 
\rightarrow \upre(\neg W(L'))$ holds.  Once $W(L) \wedge \upre(\neg W(L'))$ 
becomes unsatisfiable, i.e., no more state exists from which the environment can 
enforce to leave $W(L)$, we have found the final winning region and the algorithm 
terminates. 

\mysubsubsection{Implementation and Optimizations}
A simple realization of this approach uses a QBF solver both to compute a state 
$s$ and to generalize the induced blocking clause $\neg s$.  A generally more 
efficient approach is to use two competing SAT solvers for the two different 
quantifiers in \upre\ when computing $s$.  Other optimizations include the 
utilization of reachability information during the computation of $s$ 
and during the generalization 
of $\neg s$%
.  
A 
detailed description of different realizations and optimizations can be found 
in~\cite{BloemKS14}. 

\mysubsubsection{Extraction of Small Winning Strategies}
To obtain an implementation from the winning region, different methods can be 
applied. One possibility is to compute a certificate for the validity of the
QBF formula
$$\forall L, X_u\ \exists X_c,L'.\ W(L) \rightarrow \bigl(T(L,X_u, X_c, L') 
\wedge W(L')\bigr)$$
in the form of functions defining the variables in $X_c$ based on $L$ and $X_u$ 
using methods for QBF certification~\cite{NiemetzPLSB12}.  Another option are 
learning-based approaches that have also been proposed for BDD-based synthesis, 
but work particularly well in a SAT-/QBF-based 
framework~\cite{EhlersKH12,BloemEKKL14}.  Similar to the BDD-based methods for 
extracting small winning strategies, these learning approaches also compute 
solutions for one output $c \in X_c$ at a time.  They start with a first 
guess of a concrete output function, and then refine this guess based on 
counterexamples. 

\subsection{Template-based Synthesis}
\label{sec:templateApproach}

In order to compute a winning region, symbolically represented as a formula 
$W(L)$ over the state variables $L$, this approach constructs a parameterized 
CNF formula $\tilde{W}(L,P)$, where $P$ is a certain set of Boolean template parameters. 
Different concrete values for these parameters $P$ induce a different concrete 
CNF formula $W(L)$ over the state variables.  This is done as follows.  First, 
the approach fixes a maximum number $N$ of clauses.  Then, for every clause and 
every state variable, it introduces parameters defining whether the state variable 
occurs in the clause, whether it occurs negated or unnegated, and whether the clause is 
used at all.  This way, the search for a CNF formula over the state variables 
(the winning region $W(L)$) is reduced to a search for Boolean constants (values 
for the template parameters $P$).  A QBF solver is used to compute 
template parameter values such that (a) the winning region contains only safe 
states, (b) the winning region contains the initial state, and (c) from each 
state of the winning region, the system player can guarantee that the successor
state will also be in the winning region, regardless of the choice of the
environment. This is done by computing a 
satisfying assignment for the variables $P$ in QBF:
\begin{align*}
\exists P\ \forall L, X_u\ \exists X_c,L'.\\
 &\hspace*{-1.3cm}
\tilde{W}(L,P) \rightarrow \neg \out ~\land \\
&\hspace*{-1.3cm} \text{Init}(L) \rightarrow \tilde{W}(L,P) ~\land\\
&\hspace*{-1.3cm}\tilde{W}(L,P) \rightarrow \bigl(T(L,X_u, X_c, L')\land \tilde{W}(L',P)\bigr).
\end{align*}
More details can be found in~\cite{BloemKS14}.

%% file: problem_fig.tex
\begin{tikzpicture}[auto,
    block_center/.style ={rectangle, rounded corners, draw=black, thick, fill=white,
      text width=5em, text centered, minimum height=4em, node distance=3mm and 10mm},
    block_noborder/.style ={rectangle, draw=none, thick, fill=none,
      text width=2em, text centered, minimum height=1em, node distance=3mm and 10mm},
      arrow/.style ={draw, thick, -latex', shorten >=0pt},
    coord/.style={coordinate, node distance=6mm and 10mm}],
      \node [block_noborder] (inputs) {$X_u$};
      \node [block_center, right=of inputs, yshift=-1.3cm, minimum height=2cm] (system) {$C$};
      \node [block_center, right=of system, yshift=.5cm, minimum height=3cm] (monitor) {$M$};
      \node [block_noborder, right=of monitor, yshift=.8cm] (output) {$\out$};
    \begin{scope}[every path/.style=arrow]
      \path (inputs.east) -- ++(0.5,0) |- ([yshift=0.5cm] system.west);
      \path (inputs.east) -- ([yshift=.8cm] monitor.west);
      \path ([yshift=0.8cm] monitor.east) -- (output);
      \path ([yshift=-0.8cm] monitor.east) -| ++(0.5,-1.2) -- ++ (-3,0) |-
      ([yshift=-0.8cm] monitor.west);
      \path ([yshift=-0.8cm] monitor.east) -| ++(0.5,-1.2) -- ++ (-5.7,0) |-
      ([yshift=-0.3cm] system.west);
    \end{scope}
    \path ([yshift=0.5cm] system.east) to node {$X_c$} (monitor.west);
    \draw [arrow] ([yshift=0.5cm] system.east) -- (monitor.west);
    \path ([yshift=-0.8cm] monitor.east) to node [xshift=1em] {$L$}
    ++(0.5,-1.2);
    \path ([yshift=-0.8cm] monitor.east) to node {} ++(0.5,-1.2) to node {} ++
    (-5.7,0) to node [xshift=-1em] {$L$} ([yshift=-0.3cm] system.west);
  \end{tikzpicture}

%% file: format.tex
\section{Benchmark Format}
\label{sec:format}

For the first \syntcomp, we have chosen to use an extension of the AIGER 
format that is already used in automatic verification and is suitable for our 
selected range of problems, as well as extendable to other classes of 
problems. Furthermore, the format poses a low entry-barrier for developers of 
synthesis tools, as synthesis problems are directly given in a bit-level 
representation that can easily be encoded into BDDs and SAT-based approaches. In 
the following, we first recapitulate the AIGER format\footnote{\url{http://fmv.jku.at/aiger/}. Accessed February 2016.}, defined by 
Biere as the specification language for the hardware model checking competition 
(HWMCC)\footnote{\url{http://fmv.jku.at/hwmcc/}. Accessed February 2016.}. 
Then we show an extension of AIGER to a specification 
format for synthesis problems with safety specifications, developed for \syntcomp.
Finally, we define how to use the AIGER format
for solutions of synthesis problems in this setting.

\subsection{Original AIGER Format}

The AIGER format was developed as a compact and simple 
file format to represent benchmarks for the hardware model checking 
competition (HWMCC). Benchmarks are encoded as multi-rooted 
And-Inverter Graphs (AIGs) with latches that store the system state. We use 
version \texttt{20071012} of the format. There is an ASCII variant and a more 
compact binary variant of the format. Since the binary format is more 
restricted and thus harder to extend than the ASCII format, we have chosen 
to work with the ASCII variant for \syntcomp. In the following, we explain the 
structure of AIGER files for model checking of safety properties.

A file in AIGER format (ASCII variant) consists of the following parts:
\begin{enumerate}
\item Header,
\item Input definitions,
\item Latch definitions,
\item Output definitions,
\item AND-gate definitions,
\item Symbol table (optional), and
\item Comments (optional)
\end{enumerate}

\mysubsubsection{Header}
The header consists of a single line

\texttt{ aag M I L O A }

\noindent where \texttt{aag} is an identifier for the ASCII variant of the AIGER format,
\texttt{M} gives the maximum \emph{variable index}, and \texttt{I, L, O, A} 
the number of inputs, latches, outputs, and AND gates, respectively. 

In the rest of the specification, each input, latch, output, and AND gate is 
assigned a variable index $i$. 
To support negation, variable indices $i$ are 
even numbers, and the negation of a variable can be referred to as $i+1$. 
Variable index $0$ is reserved for the constant truth value \texttt{false}, 
and accordingly $1$ refers to \texttt{true}.
In the following, all numbers that represent inputs, latches, outputs or 
AND-gates need to be smaller or equal to $2$\texttt{M}$+1$.

\mysubsubsection{Input definitions}
Every input definition takes one line, and consists of a single number (the 
variable index of the input). 
Inputs are never directly negated, so they are always represented by even 
numbers.

\mysubsubsection{Latch definitions}
Every latch definition takes one line, and consists of an even number 
(the variable index that represents the latch), followed by a number that 
defines which variable is used to update the latch in every step. Latches 
are assumed to have initial value $0$. 

\mysubsubsection{Output definitions}
Every output definition takes one line, and consists of a single
number (representing a possibly negated input, latch, or
AND-gate).
For our class of (safety) problems, there is exactly one
output, and safety conditions are encoded such that the circuit is
safe if the output is always $0$.

\mysubsubsection{AND-gate definitions}
Every AND-gate definition takes one line, and consists of three numbers. The 
first is an even number, representing the output of the AND-gate, and is 
followed by two numbers representing its (possibly negated) inputs.

\mysubsubsection{Symbol table}
The symbol table assigns names to inputs, latches, and outputs. It is
optional, and need not be complete. Every line defines the name of one
input, latch, or output, and starts with \texttt{i,l,o}, respectively,
followed by the number of the input, latch, or output in the sequence
of definitions (\emph{not} the variable index of the input - so the
first input is always defined by a line starting with \texttt{i0}, the
first latch with \texttt{l0}). This is followed by an arbitrary string
that names the variable. 

\subsection{Modified AIGER format for synthesis specifications}
\label{sec:synthesis-input}

The \syntcomp format is a simple extension of the AIGER format for controller
synthesis: we reserve the special string ``\texttt{controllable\_}'' in
the symbol table, and prepend it to the names of controllable input
variables. All other input variables are implicitly uncontrollable.

The \emph{synthesis problem defined by an extended AIGER file} is to find a
circuit that supplies valuations for the controllable inputs, based on
valuations of uncontrollable inputs and latches 
of the given circuit, such that the output always remains $0$.

\subsection{Output of synthesis tools in AIGER format}

Starting from an input as defined in
Section~\ref{sec:synthesis-input}, we define when an AIGER file is a
\emph{solution} of the synthesis problem. Informally, the solution must contain
the specification circuit, and must be verifiable by existing model checkers
that support the AIGER format. We give a more detailed definition in the
following.

\subsubsection{Syntactic correctness}
\label{sec:AIGER-syntax}

Below we define how the input file can be changed in order to obtain a
syntactically correct solution. Unless specified otherwise below, the output
file must contain all lines of the input file, unmodified and in the same
order.

\mysubsubsection{Header} The original header line

\texttt{ aag M I L O A }

 must be modified to 

\texttt{ aag M' I' L' O A' }

where 

\begin{itemize}
\item \texttt{I'} $=$ \texttt{I} $-\ c$ 
\\(for $c$ controllable inputs in the specification)
\item \texttt{L'} $=$ \texttt{L} $+\ l$ 
\\(for $l$ additional latches defined in the controller)
\item \texttt{A'} $=$ \texttt{A} $+\ a$ 
\\(for $a$ additional AND-gates defined in the controller)
\item \texttt{M'} $=$ \texttt{I'} $+$ \texttt{L'} $+$ \texttt{A'}
\end{itemize}

The correct value for $c$ can be computed from the symbol table of the input 
file, while correct values for $l$ and $a$ depend on the number of latches 
and AND-gates in the solution.

\mysubsubsection{Inputs}
Definitions for uncontrollable inputs remain unchanged. Definitions for 
controllable inputs are removed, and the corresponding variable indices have 
to be redefined either as new latches or AND-gates (see below).

\mysubsubsection{Latches}
No definitions of latches may be removed, but additional latches may be 
defined in the lines following the original latches.

\mysubsubsection{Outputs}
No definitions of outputs may be removed, no additional outputs may be defined.

\begin{sloppypar}
\mysubsubsection{AND-gates}
No definitions of AND-gates may be removed, but additional AND-gates may be 
defined in the lines following the original AND-gates.
\end{sloppypar}

\mysubsubsection{Global restrictions} 
All variable indices of controllable inputs have to be redefined exactly 
once, either as a new latch or as a new AND-gate. New latches and AND-gates 
may be defined using the remaining (uncontrollable) inputs, any latches, or 
newly defined AND-gates, but \emph{not} original AND-gates.\footnote{The 
reason for disallowing original AND-gates is that we want the controller to 
work only based on the \emph{state} of the given circuit (i.e.,
values of latches), and the uncontrollable inputs.   
Original AND-gates can be duplicated in the controller if 
necessary.}

\mysubsubsection{Symbol table and Comments}
The symbol table remains unchanged. Comments may be removed or modified at
will.

\subsubsection{Semantic Correctness}

All input files will have the same structure as \emph{single safety property}
specifications used in HWMCC. In particular, this means that there is only one
output, and the system is safe if and only if this output remains $0$ for any
possible input sequence.

Any output file satisfying the syntactical restrictions described in
Section~\ref{sec:AIGER-syntax} is an AIGER file. 
It is \emph{correct} if for any input sequence (of the uncontrollable
inputs), the output always remains $0$. We say that it is a \emph{solution} to
the synthesis problem defined by the input file if it is successfully model
checked by an AIGER-capable model checker within a determined time bound. 

%% file: benchmarks.tex
\section{Benchmarks}
\label{sec:benchmarks}

The benchmark set for the first \syntcomp consisted of $569$ benchmark problems 
overall, out of which $390$ are realizable and $179$
unrealizable.\footnote{Numbers regarding realizability are to the best of our
  knowledge. The realizability status has not been verified for all benchmark
  instances.}  
Most of the benchmarks existed before in other formats, and have been 
translated to our new format. The full set of benchmarks used in \syntcomp 2014
is available in directory \texttt{Benchmarks2014} of our public Git repository at
\url{https://bitbucket.org/swenjacobs/syntcomp/}. In the
following, we first explain how  
benchmarks have been collected, translated and tested, and then describe the 
different sets of benchmarks.

\subsection{Collection of Benchmarks} 
One of the major challenges for the first \syntcomp was the collection of 
benchmarks in the extended AIGER format. Following the decision to use this 
format, a call for benchmarks was sent to the synthesis community. Many 
synthesis tools have their own benchmark set, but none of them previously 
used the \syntcomp format, and therefore such benchmarks had to be translated. 
Since we restrict to safety specifications currently, such a translation usually
involves a safe approximation of liveness by safety properties, and results in a
family of benchmark instances for different precision of the approximation.

\mysubsubsection{Generation and Translation of Benchmarks}
One method for obtaining benchmarks in AIGER format is based on a translation
from LTL specifications, together with a reduction to a bounded synthesis
problem, as used in Acacia+\footnote{\url{http://lit2.ulb.ac.be/acaciaplus/}. Accessed February 2016}~\cite{fjr11,bbfjr12}.
 The idea is to
\begin{inparaenum}[1)] 
\item translate the negation
of the LTL formula into a universal co-B\"uchi automaton,
\item strengthen this automaton into a universal $k$-co-B\"uchi automaton that accepts a word $w$ if and
only if all the runs on $w$ visit rejecting states at most $k$ times --- such an
automaton defines a safety objective and can be easily made deterministic.
\item Finally, a safety game is obtained by encoding succinctly this
deterministic safety automaton as an AIGER specification.
\end{inparaenum} 
We thus obtain a family
of benchmark instances, one for each valuation of $k$. If the original LTL
specification is realizable, then the resulting benchmark instance will be
realizable for sufficiently large $k$. This translation from LTL to AIGER has been implemented by Guillermo A. P\'erez in the \ltltoaig routine\footnote{\url{https://github.com/gaperez64/acacia_ltl2aig}. Accessed February 2016.}.

Another successful way of obtaining benchmarks was to 
start from Verilog code, and use a toolchain composed of the {\tt vl2mv} 
routine of the VIS system\footnote{\url{http://vlsi.colorado.edu/~vis/}. Accessed February 2016.}~\cite{VIS}, followed by translation to AIGER (and 
optimization) by ABC\footnote{\url{http://www.eecs.berkeley.edu/~alanmi/abc/}. Accessed February 2016.}~\cite{ABC}, and from binary AIGER format to ASCII format 
by the {\tt aigtoaig} routine from the AIGER tool set\footnote{\url{http://fmv.jku.at/aiger/}. Accessed February 2016.}. Liveness
properties can be approximated by safety properties, and we obtain a family of
benchmark instances for different approximations. Such an approximation is
explained in more detail in Section~\ref{sec:bench-genbuf}. This approach will
be called the \emph{Verilog toolchain} below.

Finally, a number of benchmarks have been obtained by translation from
structured specifications for the generalized reactivity(1) game solver
SLUGS\footnote{\url{http://github.com/ltlmop/slugs}. Accessed February 2016.}. The term ``structured'' in this context refers to support for 
 constraints over (non-negative) integer numbers, which are automatically
 compiled into a purely Boolean form.
The purely Boolean generalized reactivity(1) safety specification is then
translated into a monitor automaton in AIGER format, which is finally optimized
using the ABC toolset by applying the command sequence {\tt
  rewrite}. We will call this approach the \emph{SLUGS toolchain} below.

\mysubsubsection{Testing and Classification of Benchmarks}
To test the resulting benchmarks, we fed them to our reference 
implementation Aisy\footnote{\url{https://bitbucket.org/art_haali/aisy-classroom}. Accessed February 2016.},
 and compared the produced solution to the expected result.
Since our reference implementation is not as efficient as the participants of
the competition, a significant number of benchmarks was only solved during the
competition, but not in our initial tests. Those that were not solved were
classified into realizable and unrealizable according to informed guesses of the
benchmark authors. During the competition, this resulted in $3$ problem
instances being re-classified from unrealizable to realizable, or vice versa.

\subsection{Toy Examples} 
These benchmarks are based on original 
Verilog specifications that have been translated to AIGER using the Verilog
toolchain. The set includes
specifications of basic circuits like adders, bit shifters, multipliers, and 
counters. Additionally, it contains some specifications with typically 
very simple properties, e.g., that outputs must match inputs, or that the XOR 
of inputs and outputs must satisfy some property. All examples are 
parameterized in the bit-width of the controllable and uncontrollable inputs, 
ranging between $2$ and $128$ bits on some examples, and for each example 
there are two versions, using the optimizing and non-optimizing translation by 
ABC, respectively. Overall, this set contains $138$ benchmarks.

All AIGER files contain the original Verilog code, as well as the commands 
used to produce the AIGER file, in the comments section. This set of benchmarks
was provided by Robert K\"onighofer.

\subsection{Generalized Buffer} 
\label{sec:bench-genbuf}
The well-known specification of an industrial generalized buffer was 
developed by IBM and subsequently used as a synthesis benchmark for 
Anzu~\cite{Anzu} and other tools. It is parameterized by the number of 
senders which send data to two receivers. The buffer has a handshake protocol 
with each sender and each receiver. A complete Genbuf consists of a 
controller, a FIFO, and a multiplexer. In this benchmark, the FIFO 
and multiplexer are considered as part of the environment, and the controller is
synthesized. As a synthesis case study for Anzu, it has been explained in 
detail by Bloem et al.~\cite{BloemGJPPW07}. Robert K\"onighofer translated these
benchmarks to AIGER, as explained in the following.

\paragraph{Liveness-to-Safety Translation.}
For Anzu, the Genbuf benchmark contains B\"uchi assumptions $\{A_1, \ldots, 
A_m\}$ that are satisfied if all state sets $A_i$ are visited infinitely often, 
and B\"uchi guarantees $\{G_1, \ldots, G_n\}$ requiring that all $G_j$ are 
visited infinitely often if all assumptions are satisfied.  Three different 
translations into safety specifications were performed.  Translation 
``\texttt{c}'' (for \emph{counting}) applies the well-known counting construction:  A modular counter 
$i\in\{0, \ldots, m\}$ stores the index of the next assumption.  If an 
accepting state $s \in A_i$ of this next assumption is visited, the counter is 
incremented modulo $m+1$.  If $i$ has the special value $0$, it is always 
incremented.  The same counting construction is applied to the B\"uchi 
guarantees with counter $j\in\{0,\ldots, n\}$. Finally, a third counter $r$ 
is used to enforce a minimum ratio between the progress in satisfying 
guarantees and assumptions:  Whenever $j$ is incremented, $r$ is reset to $0$. 
Otherwise, if $i=0$, then $r$ is incremented.  If $r$ ever exceeds some bound 
$k$, then \out is set.  A controller enforcing that \out cannot become $1$ thus 
also enforces that all $G_j$ are visited infinitely often if all $A_i$ are 
visited infinitely often.  Translation ``\texttt{b}'' (for \emph{bitwise}) is similar but uses one 
bit per assumption and guarantee instead of a counter.  It thus avoids imposing 
an (artificial) order between properties.  Translation ``\texttt{f}'' (for \emph{full set}) is 
similar to ``\texttt{b}'' but resets $r$ only if all guarantees have been seen 
in a row (rather than only the next one).

\paragraph{Translation to AIGER.}
Anzu comes with scripts to construct Genbuf benchmark instances with different 
numbers of senders.  These scripts were modified to output a Verilog 
representation, and from there the Verilog tool\-chain was used to obtain 
benchmarks in AIGER format.  The final specification is parameterized in 
\begin{inparaenum}[1)] \item the number of senders which send data, \item the 
type (\texttt{c}, \texttt{b} or \texttt{f}) and the bound $k$ 
of the liveness-to-safety translation, and \item whether or not ABC 
optimizations are used in the translation. \end{inparaenum} All AIGER files 
contain the original Verilog code (which in turn contains the Anzu specification 
it was created from), as well as the commands used to produce the AIGER file, in 
the comments section. Overall, this set contains $192$ benchmark instances.

\subsection{AMBA Bus Controller}
This is a specification of an arbiter for the AMBA AHB bus, based on an 
industrial specification by ARM. Like the Genbuf case study, it has 
been used as a synthesis benchmark for Anzu~\cite{Anzu} and other tools. It 
is parameterized with the number of masters that can access the bus and have 
to be coordinated by the arbiter. The AMBA AHB bus allows masters to request 
different kinds of bus accesses, either as a single transfer or as a burst, 
where a burst can consist of either a specified or an unspecified number
of transfers. Besides correct modeling of these different forms of accesses, 
the specification requires responses to all requests (that are not eventually 
lowered), as well as mutual exclusion of bus accesses by different masters.
As a synthesis case study for Anzu, it has been explained in 
detail by Bloem et al.~\cite{BloemGJPPW07a}. 

\begin{sloppypar}
The Anzu specification has been translated by Robert K\"onighofer to AIGER 
format using the Verilog toolchain in the same way as for the Genbuf benchmark. 
 Instances are parameterized in the number of masters, 
and (as for Genbuf) the type (\texttt{c}, \texttt{b} or \texttt{f}) and the 
bound $k$ of the liveness-to-safety translation, as well as 
whether or not ABC optimizations are used in the translation.
All AIGER files contain the original Verilog code (which in turn contains the 
Anzu specification it was created from), as well as the commands 
used to produce the AIGER file, in the comments section. Overall, this set 
contains $108$ benchmarks.
\end{sloppypar}

\input{ULBbenchmarks}

\subsection{Factory Assembly Line}
This benchmark models an assembly line with multiple tasks that need to be 
performed on the objects on the conveyor belt. The specification models a number
of robot arms (fixed to $2$), a number $n$ of objects on the conveyor belt, and
a number $m$ of tasks that may have to be performed on each object before it
leaves the area that is reachable by the arms. The belt moves 
after a fixed number $k$ of time steps, pushing all objects forward by one place,
and the first object moves out of reach of the arms (while a new object enters
at the other end of the belt). The arms are modeled such that they cannot occupy
the space above the same object on the belt, and can move by at most one
position per time step. In particular, this means that they cannot pass each
other. Whenever an arm is in the same position as an object that has unfinished
tasks, it can perform one task on the object in one time step. Usually, the
assumption is that at most $m-1$ of the $m$ tasks need to be performed on any
object, but there may be a fixed number $c$ of \emph{glitches} in an execution
of the system, which means that an object with $m$ open tasks is pushed onto the
belt.

This specification has been translated by R\"udiger Ehlers from original
benchmarks for the SLUGS GR(1) synthesis 
tool, using the SLUGS toolchain. Overall, this set
contains $15$ benchmarks, for different values of $m$ ($3$ to $7$), $n$ ($3$ to
$6$), $k$ ($1$ to $2$) and $c$ ($0$ to $11$).

\subsection{Moving Obstacle Evasion}
This benchmark models a controller for a robot that moves on a quadratic grid of
parametric size $m$, and has to avoid colliding with a moving obstacle (of size $2
\times 2$ grids). In any time step, the robot and the obstacle can only move by
at most one grid in $x$ and $y$ direction. Additionally, the obstacle can
usually only move at most every second time step. However, as in the assembly
line benchmarks, there may be a fixed number $c$ of glitches in an execution of
the system, which in this case means that the obstacle moves even though it has
already moved in the immediately preceding time step.

This specification has been translated by R\"udiger Ehlers from original
benchmarks for the SLUGS GR(1) synthesis 
tool, using the SLUGS toolchain. Overall, this set
contains $16$ benchmarks, for different values of $m$ ($8$ to $128$) and $c$
($0$ to $60$).

%% file: ULBbenchmarks.tex
\subsection{LTL2AIG Benchmarks}
This set contains several benchmarks provided in the Acacia+ tool
package~\cite{bbfjr12}, translated into AIGER format
using the \ltltoaig routine. 
The set includes:
\begin{itemize}
\item $50$ benchmarks from the test suite included with the synthesis tool
\textsc{Lily}\footnote{\url{http://www.iaik.tugraz.at/content/research/opensource/lily/}. Accessed February 2016.}~\cite{JobstmannB06}, with specifications of traffic lights
and
arbiters in different complexity ($25$ original examples, each with $2$
different choices of $k$).
\item $4$ versions of the Genbuf case study, but in a much more challenging form
  than the specification mentioned in Section~\ref{sec:bench-genbuf}\footnote{We
    conjecture that this version is more challenging because it is based on a
    large LTL specification, which is translated to a single, very big B\"uchi
    automaton in the first step of the \ltltoaig routine. This results in a
    circuit that is much more complex than the ones from
    Section~\ref{sec:bench-genbuf}.}  This
  version is only specified for $2$ senders and $2$
receivers, and for $4$ different choices of $k$.
\item $5$ versions of a load balancer case study, originally presented with
synthesis tool \textsc{Unbeast}\footnote{\url{http://www.react.uni-saarland.de/tools/unbeast/}. Accessed February 2016.}~\cite{Ehlers11}.
\item $23$ benchmarks that use the synthesis tool to obtain a deterministic 
B\"uchi automaton for the given LTL specification (if possible), and
\item $18$ benchmarks for a similar conversion from LTL to deterministic
parity automata. The latter two conversions are mentioned as applications of
synthesis procedures by Kupferman and Vardi~\cite{KupfermanV05}.
\end{itemize}
Overall, this set contains $100$ benchmarks.

%% file: rules.tex
\section{Rules}
\label{sec:rules}

The rules for \syntcomp were inspired by similar competitions such as the SAT 
competition and the HWMCC. The basic idea is that submitted tools are 
evaluated on a previously unknown set of benchmarks, without user 
intervention. A simple ranking of tools can be obtained by checking only the 
correctness of solutions, and counting the number of problem instances that 
can be solved within a given timeout. However, the goal of synthesis is to 
obtain implementations that are not only correct, but also efficient. 
Therefore, we also considered refined rankings based on the \emph{quality} 
of the produced solutions, measured by the size of the implementation.

\mysubsubsection{Tracks} The competition was separated into two tracks: the \emph{realizability 
track} which only required a binary answer to the question whether or not 
there exists a circuit that satisfies the given specification, and the \emph{
synthesis track} which was only run on realizable benchmarks, and asked for a 
circuit that implements the given specification. The motivation for this 
split was again to have a low entry-barrier for tool creators, as an 
efficient realizability checker can be implemented with less 
effort than a full synthesis tool that produces solutions and optimizes them 
for size. Indeed, $2$ out of $5$ submitted tools make use of this split and 
only supply a realizability checker, and these two tools solve 
more problems in the realizability track than any of the full synthesis tools. 

\mysubsubsection{Subtracks} Each track was divided into a \emph{sequential} subtrack, where 
tools can use only one core of the CPU, and a \emph{parallel} subtrack, 
allowing tools to use multiple cores in parallel. The decision to have both 
sequential and parallel execution modes was based on the expectation that 
parallelization would often be trivial---i.e., a number of different but 
largely independent strategies running in parallel.\footnote{In particular, 
non-trivial parallelization is difficult for BDD-based tools, since none of 
the 
existing parallel BDD packages supports all features needed for the 
optimizations mentioned in Section~\ref{sec:BDDgame}.} Therefore, we also wanted to 
evaluate tools in sequential execution mode in order to measure and identify 
the single best 
strategy.

\subsection{Entrants}
We asked for synthesis tools to be supplied in source code, licensed
for research purposes, and we offered to discuss possible solutions if this
restriction was a problem to any prospective participant. This was not the case
for any of the research groups that contacted us regarding the competition. 
The organizers reserved the right to 
submit their own tools and did so in the form of \basil, implemented 
by R. Ehlers, and \demiurge, implemented in part in the research group of R.
Bloem. We encouraged participants to visit SYNT and CAV for
the presentation of the \syntcomp results, but this was not a requirement for
participation. 

We allowed up to $3$ submissions per author and subtrack, where submissions 
are considered to be different if source code, compilation options or command 
line arguments are different. This limit was 
chosen to allow some flexibility for the tool creators, while avoiding the 
flooding of the competition with too many different configurations of the 
same tool. All tools must support the input and output format of \syntcomp, 
as defined in Section~\ref{sec:format}. Additionally, each entrant to 
\syntcomp was required to include a short system description.

The organizers commited to making reasonable efforts to install each tool but 
reserved the right to reject entrants where installation problems persisted. 
This was not the case for any of the entrants. Furthermore, in case of 
crashes or wrong results we allowed submission of bugfixes if possible within 
time limits. In one case, a bugfix was submitted that resolved a number of 
solver crashes that only appeared during the competition runs.

\subsection{Ranking}
\label{sec:rankings}

In both the realizability and the synthesis track, competition entrants were 
ranked with respect to the number of solved problems. Additionally, we 
consider a more fine-grained \emph{relative ranking} that distributes points 
for each benchmark according to the relative performance of tools, measured either in
the time needed to find a solution, or the size of 
the solution. A drawback of this relative 
ranking is that it does not allow easy comparison to tools that did not 
participate. As an alternative that resolves this problem, we additionally 
give a \emph{quality ranking} for the synthesis track that compares the size 
of the provided solution to a reference size.\footnote{The quality ranking was 
devised for the second \syntcomp and 
was applied to the results of the first competition only after the 
presentation of 
results at SYNT and CAV 2014.}

For all rankings, a timeout (or no answer) gives $0$ points. A punishment for 
wrong answers was not necessary, since the full set of benchmarks was made 
available to the participants one month 
before the submission of solvers. 

\mysubsubsection{Correctness and Ranking in Realizability Track} 
For the 
realizability track, the organizers and benchmark authors took responsibility 
for determining in advance whether specifications are realizable or 
unrealizable, by using knowledge about how the benchmarks were generated. 
When in doubt, a majority vote between all tools that solved a given 
benchmark was used to determine the correct outcome.\footnote{This rule only 
had to be used in one instance, where a benchmark was solved by only one 
tool, and was reported to be realizable although unrealizable was the 
expected outcome. In our analysis it turned out that the tool was correct, 
and the initial classification as unrealizable was wrong.} 

In addition to a 
ranking based on the number of solved problem instances, tools were evaluated 
with a relative ranking based on the time needed to come to the solution, 
where the tool with the smallest time earns the highest rank (see below). For 
the sequential subtrack, tools were ranked with respect to CPU time, 
while for the parallel subtrack we ranked tools with respect to wall-clock time.

\mysubsubsection{Correctness and Ranking in Synthesis Track} In the synthesis 
track, correctness of solutions was assessed by checking both syntactical 
and semantical correctness. Syntactical correctness means conformance to our 
output format defined in Section~\ref{sec:format}, which was checked by a 
separate syntax checker. Semantical correctness was tested by a model checker 
({\tt iimc}\footnote{\url{http://ecee.colorado.edu/wpmu/iimc/}. Accessed February 2016.}, based on the IC3 
algorithm~\cite{Bradley11}), which had to terminate within a separate time 
bound for the result to be considered correct. As in the realizability 
track, there is a ranking with respect to the number of solved problem 
instances, as well as a relative ranking. The latter is in this case based on 
the size of solutions, given by the number of AND-gates in the resulting 
circuit. In addition, we provide a quality ranking that awards points for 
every solution, based on a comparison of the solution size to a reference size (see below).

\mysubsubsection{Relative Ranking} 
For every benchmark, all tools that 
provide a correct solution are ranked with respect to the metric 
(time or size), 
and each tool obtains points based on its rank. 
In detail: if $k$ benchmarks are used in the track, then $1000/k = p$ points are awarded per benchmark. 
If $n$ tools solve the benchmark, then the points for 
that benchmark are divided into $\Sigma_{i=1}^n i = f$ fractions, and the tool 
which is at rank $m$ will get $\frac{n-m+1}{f}\cdot p$ points for this 
benchmark.

\mysubsubsection{Quality Ranking}
In the \emph{quality ranking}, solutions are awarded 
points depending on the size $size_{\mathit{new}}$ of the solution and a reference 
size $size_{\mathit{ref}}$. The number of points for a solution is 
$$2 - \log_{10} \left( \frac{size_{\mathit{new}}}{size_{\mathit{ref}}}\right).$$
That is, a solution that is of size $size_{\mathit{ref}}$ gets $2$ points; a solution 
that is bigger by a factor of $10$ gets $1$ point; a solution that is bigger 
by a factor of $100$ (or more) gets $0$ points; and similarly for solutions 
that are smaller than $size_{\mathit{ref}}$.

Since for the first competition we do not have reference solutions for any 
of the problem instances, we use the smallest size of any of the solutions of 
this competition as the reference size. In future competitions, or for 
comparison of tools that did not participate, the size of the smallest 
solution that has been provided in any of the competitions before can be used.

%% file: tools.tex
\section{Participants}
\label{sec:participants}

Five systems were entered into the first \syntcomp.
In the following, we give a brief description of the methods implemented in
each of these systems. For the BDD-based tools, Table~\ref{tab:optimizations}
shows which of the optimizations from Section~\ref{sec:BDDgame} are implemented
in which tool.

\begin{table*}
\caption{Optimizations implemented in BDD-based Tools.}
\label{tab:optimizations}
\centering
\begin{tabular}{r|ccccc}
  Technique                       & \abssynthe & \basil & \realizer & \simpleBDD \\  \hline
automatic reordering                        & x          & x      & x         & x \\
eager deallocation of BDDs       &            & x      &           & x \\
direct substitution               & x          & (x)      & x         & x \\
partitioned transition relation   & x          &        & x         & x \\
simultaneous conjunction and abstraction & &  & & x \\
co-factor based extraction of winning strategies & x & x & N/A      & N/A\\
forward reachability analysis     & x          & x      & N/A       & N/A \\
\Abc minimization                 &            & x      & N/A       & N/A\\
additional optimizations (see tool descriptions)         & x          & x      & x         & x \\
\end{tabular}
\end{table*}

\subsection{\abssynthe: an abstract synthesis tool}
\label{sec:abssynthe}

\abssynthe was submitted by R. Brenguier, G. A. P\'erez, J.-F. Raskin, and 
O. Sankur from Universit\'e Libre de Bruxelles. \abssynthe implements a 
BDD-based synthesis approach and competed in all subtracks.

\mysubsubsection{Synthesis algorithms}
\abssynthe implements different BDD-based synthesis algorithms, with and without
abstraction, described in more detail in~\cite{BrenguierPRS14}. All algorithms
use the BDD package CUDD~\cite{somenzi99}, with automatic BDD reordering using the sifting
heuristic.

The \emph{concrete algorithm with partitioned transition relation} (C-TL)
implements BDD-based synthesis with partitioned transition relation and direct
substitution of state variables with BDDs. In addition, when computing
$\upre(S(L'))$, then the transition functions $f_l$ of all latches are first
restricted to $\neg S(L)$, effectively only computing the uncontrollable
predecessors which are not already in $S(L)$. These new states are then joined
to $S(L)$, which gives the same result as the standard $\upre$ computation. 

The \emph{basic abstract algorithm} (A) implements synthesis
with a precomputed (monolithic) abstract transition relation, and some additional
optimizations. 

The \emph{alternative abstract algorithm} (A-TL) avoids using a precomputed
transition relation by implementing abstract operators for post-state
computation.

\abssynthe was intended to compete in these different configurations. However,
due to a miscommunication between tool authors and competition organizers, the
necessary command line parameters were not used, such that only one
configuration participated, namely (C-TL). 
Unfortunately, this error
was discovered too late to run the additional configurations before the 
presentation of results at CAV 2014. 

However, as mentioned
in~\cite{BrenguierPRS14}, the abstraction-based methods overall performed
worse than the concrete algorithm (C-TL), and thus the fastest configuration did 
participate in the competition. 


\mysubsubsection{Strategy extraction} 
Strategy extraction in \abssynthe uses the co-factor-based approach described in
  Section~\ref{sec:BDDgame}, including the additional forward reachability
  check. When extracting the circuit, every AIG node constructed from the BDD
  representation is cached in order to avoid duplicating parts of the circuit.




\mysubsubsection{Implementation, availability}
\abssynthe is written mostly in Python, and depends only on a simple AIG
library (fetched from the AIGER toolbox\footnote{\url{http://fmv.jku.at/aiger/}. Accessed February 2016.}) and the  BDD
package CUDD\footnote{\url{http://vlsi.colorado.edu/~fabio/CUDD/}. Accessed February 2016.}. The source code is available at
\url{https://github.com/gaperez64/AbsSynthe}.

\subsection{\basil: \textbf{B}DD-b\textbf{a}sed \textbf{s}afety synthes\textbf{i}s too\textbf{l}}
\basil was submitted by R. Ehlers from the University of Bremen, and
implements a BDD-based synthesis approach. \basil competed in all subtracks.

\mysubsubsection{Synthesis algorithm}
\basil implements a BDD-based synthesis algorithm, based on the BDD package
CUDD. It uses automatic reordering of BDDs with the \emph{sifting} heuristic,
reconfigured in order to optimize more greedily. In
contrast to all other BDD-based tools in the competition, it does not use a partitioned transition relation. It does however use a technique similar to direct
substitution, regarding latches that are always updated by the value of an input
variable: BDD variables that represent such inputs are double-booked as both
an input and a post-state variable of the latch, and therefore need not be
explicitly encoded into the transition relation.
Additionally, when building the transition relation
it eagerly deletes BDDs that are only used as intermediate values as soon as they
are not needed anymore. This is the case if a gate $A$ is not used as a
controllable input or an input to a latch, and all nodes that depend on $A$ have
been processed.

\mysubsubsection{Strategy extraction}
\basil computes strategies with the co-factor-based approach from
Section~\ref{sec:BDDgame}, including forward reachability analysis and \Abc
minimization.

As an additional optimization, during strategy extraction the output bit BDDs are reduced
  in size by applying \textit{LICompaction} \cite{HongBBM00}: A joint BDD for
  all output bit BDDs is built and then, in a round-robin fashion over the
    outputs, the size of 
  the joint BDD is reduced by changing a part of it that describes the behavior
  of a single output bit in a way that makes the overall BDD smaller, but yields
  behavior that is contained in the most general strategy for winning the
  game. In order to minimize the care set for this operation, a reachable-state
  computation is performed before every step. When no further size reduction is
  found to be possible, or some timeout has been reached, optimization by
  \textit{LICompaction} is aborted. 

\mysubsubsection{Implementation, availability}
\basil is implemented in C++ and depends on the BDD package CUDD, as well as
(optionally) \Abc\footnote{\url{http://www.eecs.berkeley.edu/~alanmi/abc/}. Accessed February 2016.} for strategy minimization. It is currently not publicly available.

\input{demiurge}

\input{realizer}

\input{simpleBDD}

%% file: demiurge.tex
\subsection{Demiurge}

\demiurge was submitted by R. K\"onighofer from Graz University of 
Technology and M. Seidl from Johannes-Kepler-University Linz. \demiurge 
implements incremental SAT- and QBF-based synthesis as described in
Section~\ref{sec:SATgame}, as well as  
template-based synthesis with QBF solving as described in
Section~\ref{sec:templateApproach}. Demiurge competed in all 
subtracks.

\mysubsubsection{Synthesis algorithms}
\demiurge implements different synthesis algorithms in different back-ends,
described in more detail in~\cite{BloemKS14}.

\begin{sloppypar}
The \emph{learning-based back-end} uses the incremental synthesis approach to 
compute a winning region based on two competing SAT solvers to compute and
generalize states to be removed from the 
winning region (algorithm \textsc{LearnSat} from~\cite{BloemKS14} with 
optimization \textsc{RG} enabled, but optimization
\textsc{RC} disabled). \minisat version 2.2.0 is used as underlying SAT 
solver.
\end{sloppypar}

The \emph{parallel back-end} implements the same 
method with three threads refining the winning region in parallel. 
Two threads perform the work of the learning-based back-end, one using \minisat
2.2.0 and the other using \lingeling 
\textsf{ats}. The third thread generalizes existing clauses of the winning 
region further by trying to drop more literals. Using different solvers in the 
threads is beneficial because the solvers can complement each other, sometimes 
yielding a super-linear speedup~\cite{BloemKS14}. 

The \emph{template-based back-end} uses a QBF solver to compute a winning region as 
instantiation of a template for a CNF formula over the state variables. 
For \syntcomp, \depqbf 3.02 is used as QBF solver via its API.  \bloqqer,
extended to preserve satisfying assignments~\cite{SeidlK14}, is used as QBF 
preprocessor. 

\demiurge contains more back-ends that are either experimental or did not turn 
out to be particularly competitive, and therefore did not enter the 
competition. This includes a re-implementation of the technique of Morgenstern et 
al.~\cite{MorgensternGS13}, and an approach 
based on a reduction to Effectively Propositional Logic (EPR). Details can be 
found in~\cite{BloemKS14}.

\mysubsubsection{Strategy extraction}
\demiurge provides several methods for computing strategies from the winning
region.  The algorithm used in the competition uses a computational learning
approach as proposed in~\cite{EhlersKH12}, but implemented with 
incremental SAT solving or incremental QBF solving instead of BDDs.
In terms of~\cite{BloemEKKL14}, it uses the SAT-based learning method without
the dependency optimization, with \lingeling \textsf{ats} as SAT solver.  
\Abc minimization is used in a post-processing step. 

\mysubsubsection{Implementation, availability}
\demiurge is implemented in C++, and depends on a number of underlying
reasoning engines, some of them mentioned above.
Because of its modular architecture, 
\demiurge is easily extendable with new 
algorithms and optimizations (cf.~\cite{BloemEKKL14}), thus providing a framework
for implementing new synthesis algorithms and reducing the entry barrier
for new research on SAT- and QBF-based synthesis algorithms and optimizations.
\begin{sloppypar}
\demiurge is available under the GNU LGPL license (version 3) at
\url{http://www.iaik.tugraz.at/content/research/design_verification/demiurge/}.
\end{sloppypar}

%

%% file: realizer.tex
\subsection{REALIZER-- CUDD Based Safety Game Solver}

\begin{sloppypar}
\realizer was submitted by L. Tentrup from Saarland University, 
Saarbr\"ucken. \realizer implements BDD-based realizability checking, and
competed in both realizability subtracks.
It does not support extraction of strategies.
\end{sloppypar}

\mysubsubsection{Synthesis algorithms}
\realizer is based on BDD package CUDD, and uses automatic reordering of BDDs
with the \emph{lazy sift} reordering scheme. When building the BDDs that
represent the transition relation, it uses a temporary hash table to save the
BDDs for AND gates in the AIG.
Before starting the fix-point algorithm, it builds the basic data structures used
in the fixed point calculation, like the arrays mapping the current state
variable to the next (primed) state variable or the BDD cubes used for the
existential and universal abstraction.

The actual fix-point algorithm is implemented in two variants, differing only in
the way they handle the forced predecessor function: one variant uses a
monolithic transition relation, while the other uses a partitioned transition
relation. Both variants use direct substitution.

The \emph{variant with partitioned transition relation} overall performed better
in preliminary experiments, so only this one was entered into the competition in
the sequential realizability track. Since on some examples the other variant
performed better, the \emph{parallel version} uses both variants running
(independently) in parallel.\footnote{Analysis of results and subsequent
  inspection of the source code by the tool author showed that due to a bug the
  parallel version did not work as intended, and instead used two threads with
  identical strategy. As can be seen in the results section, this lead to a
  decreased performance overall.}




\mysubsubsection{Implementation, availability}
\realizer is written in Python and uses the BDD library CUDD in version 2.4.2
with the corresponding Python bindings PyCUDD in version 2.0.2. It is currently not publicly available.

%% file: simpleBDD.tex
\subsection{Simple BDD Solver}

Simple BDD Solver was submitted by L. Ryzhyk from NICTA, Sydney and the 
Carnegie Mellon University, Pittsburgh, and A. Walker from NICTA, Sydney. Simple
BDD Solver implements BDD-based realizability checking, and
only competed in the sequential realizability subtrack.
It does not support extraction of strategies.

\mysubsubsection{Synthesis algorithm(s)}
Simple BDD Solver is a substantial simplification of the solver that was
developed for the Termite project\footnote{\url{http://termite2.org}. Accessed February 2016.}, adapted to safety
games given in the AIGER format.
It uses the BDD package CUDD, with dynamic variable reordering using the sifting
algorithm, and eager deallocation of BDDs. Furthermore, it uses
a partitioned transition relation, direct substitution, and simultaneous
conjunction and abstraction.

Additionally, it uses an alternative form for the fixpoint computation of
$\upre$ that avoids creating the $\out$ latch and simplifies the quantification structure:

The fixpoint formula (\ref{eq:fixpoint}) in Section~\ref{sec:BDDgame} is
equivalent to
\begin{align*}
\mu S(L).\ \exists X_u \forall X_c \exists L'.\\
 &\hspace*{-2.3cm}
\bigl( S(L') \land T(L, X_u, X_c, L') \bigr) \lor \out'.
\end{align*}
To avoid introducing the latch for $\out$, we substitute $\out'$ with the update
function for $\out$ - an expression $\neg SAFE(L, X_u, X_c)$ over latches and
inputs. This results in:

\begin{align*}
\mu S(L).\ \exists X_u \forall X_c \exists L'. \\
 &\hspace*{-2.3cm}
\bigl( S(L') \land T(L, X_u, X_c, L') \bigr) \lor \neg SAFE(L, X_u, X_c).
\end{align*}

Then, quantifiers are re-arranged to

\begin{align*}
\mu S(L).\ \exists X_u \forall X_c. \\
 &\hspace*{-2.3cm}
\bigl( \exists L'.\ S(L') \land T(L, X_u, X_c, L') \bigr) \lor \neg SAFE(L, X_u, X_c),
\end{align*}

with the safety condition outside of the innermost existential quantification.
With this formula for the fixpoint, simultaneous conjunction and abstraction
can be used on the left-hand side of the disjunction, and we avoid to build the
potentially large BDD of the conjunction in the left-hand side at every
iteration.

Furthermore, the tool implements a variant of the fixpoint algorithm with an
abstraction-refinement loop inspired by~\cite{dealfaro}. Since this variant has
not been found to be competitive on the set of competition benchmarks, it has
not been entered into the competition.

\mysubsubsection{Implementation, availability}
Simple BDD solver is written in the Haskell functional programming language. It
uses the CUDD package for BDD manipulation and the
Attoparsec\footnote{\url{https://hackage.haskell.org/package/attoparsec}. Accessed February 2016.} Haskell package for fast parsing. Altogether, the
solver, AIGER parser, compiler and command line argument parser are just over
300 lines of code. The code is available online at:
\url{https://github.com/adamwalker/syntcomp}.

%% file: execution.tex
\section{Execution}
\label{sec:execution}

\syntcomp 2014 used a compute cluster of identical machines with octo-core 
Intel Xeon processors (2.0 GHz) and 64 GB RAM, generously provided by Graz 
University of Technology. The machines are running a GNU/Linux system, and 
submitted solvers were compiled using GCC version 4.7. Each node has a local 
400 GB hard drive that can be used as temporary storage.

\begin{sloppypar}
The competition was organized on the EDACC platform~\cite{BalintDGGKR11} 
developed for the SAT Competitions~\cite{JarvisaloBRS12}. EDACC directly 
supports the definition of subtracks 
with different benchmark sets, different solver configurations, verification 
of outputs, and automatic distribution of jobs to compute nodes.
 During the competition, a complete node 
was reserved for each job, i.e., one synthesis tool (configuration) running 
one benchmark. This ensures a very high comparability and 
reproducibility of our results. Olivier Roussel's 
\texttt{runsolver}~\cite{Roussel11}
was used to run each job and to measure CPU time and Wall time, as well as 
enforcing timeouts. As all nodes are 
identical and no other tasks were run in parallel, no other limits than a 
timeout per benchmark (CPU time for sequential subtracks, wall time for
parallel subtracks) was set. 
The timeout for each task in any subtrack 
was $5000$ seconds (CPU time or wall time, respectively).
The queueing system in use is 
TORQUE\footnote{\url{http://www.adaptivecomputing.com/products/open-source/torque/}. Accessed February 2016.}.
\end{sloppypar}

Some solvers did not conform completely with the output format specified by 
the competition, e.g. because extra information was displayed in addition to 
the specified output. For these solvers, small wrapper scripts were used to 
execute them, filtering the outputs as to conform to the specified format.

\mysubsubsection{Validity of results}
Beyer et al.~\cite{BeyerLW15a} recently noted that \texttt{runsolver}, along 
with a 
number of other benchmarking tools, has certain deficits that endanger the 
validity of results.
In particular, the CPU time of child processes may not be measured correctly. 
First, note that CPU time is only relevant for our 
results in the sequential subtracks, where tools are restricted to a 
single CPU core. Furthermore, for the participants of \syntcomp we note that 
the only child processes (if any) are the reasoning engines for BDDs and SAT 
or QBF formulas. Since these reasoning engines take up almost all of the CPU 
time in solving synthesis tasks, a comparison of CPU time to the recorded wall 
time would most probably reveal measurements that exclude child processes. This was not 
the case for our results.

%% file: results.tex
\section{Experimental Results and Analysis}
\label{sec:results}

\begin{sloppypar}
We present the results of \syntcomp 2014, separated into realizability and
synthesis tracks, followed by some observations on the state of the art. $5$
tools entered the competition and ran in $8$ different configurations in the $4$
tracks of the competition.\footnote{As mentioned
  in Section~\ref{sec:abssynthe}, \abssynthe was supposed to compete in
  different configurations, but due to a
  miscommunication was always started in the same configuration. The results
  presented here for the relative ranking differ from those presented at CAV 2014 in that
  only one of the three identical configurations of \abssynthe is counted in
  the sequential tracks.}
All of
the results can be viewed online in our EDACC system at
\url{https://syntcompdb.iaik.tugraz.at/2014/experiments/}. Furthermore, the full experimental data, including problem instances, executable code of the solvers, logfiles of executions, solutions produced by solvers, and executable code for verifying the solutions is available in directory \texttt{ExperimentalData2014} of our public Git repository at
\url{https://bitbucket.org/swenjacobs/syntcomp/}.
\end{sloppypar}

\subsection{Realizability Track}
In the realizability track, tools were run on the full set of $569$ 
benchmarks. All $5$ tools that entered \syntcomp competed with at least one
configuration in the sequential subtrack, and $4$ of them also competed in
the parallel subtrack.

\begin{sloppypar}
\mysubsubsection{Sequential Subtrack}
The sequential realizability track had $6$ participants: \abssynthe, \basil, \realizer and
\simpleBDD competed with one configuration each, whereas
\demiurge competed with two different configurations. 
Table~\ref{tab:RealSeq} contains the number of instances
solved within the timeout per tool, the number of instances solved uniquely by a solver, and the accumulated points per tool according to our relative ranking scheme.

No tool could solve all $569$ benchmarks, and
$13$ benchmarks were not solved by any of the tools 
within the timeout.
$12$ benchmarks were solved uniquely by one
tool:
\begin{itemize}
\item \basil: $4$ versions of the factory 
assembly benchmarks (size 5x5 and 7x5, each with 10 and 11 errors)
\item \realizer: \texttt{gb\_s2\_r2\_1\_UNREAL}\footnote{This benchmark was found
    to be realizable by the tool, although it was classified as unrealizable by
    the benchmark authors. Our analysis confirmed it to be realizable.}
\item \demiurge \templ: \texttt{mult1x} with $\texttt{x} \in \{ 2,3,4,5,6 \}$,\\
  \texttt{stay18n} and \texttt{stay20n}
\end{itemize}
Table~\ref{tab:unsolved} gives an overview of benchmark instances that were
solved uniquely or not solved at all, for all subtracks of the competition.
\end{sloppypar}

\begin{figure*}
\centering
\includegraphics[width=.95\linewidth]{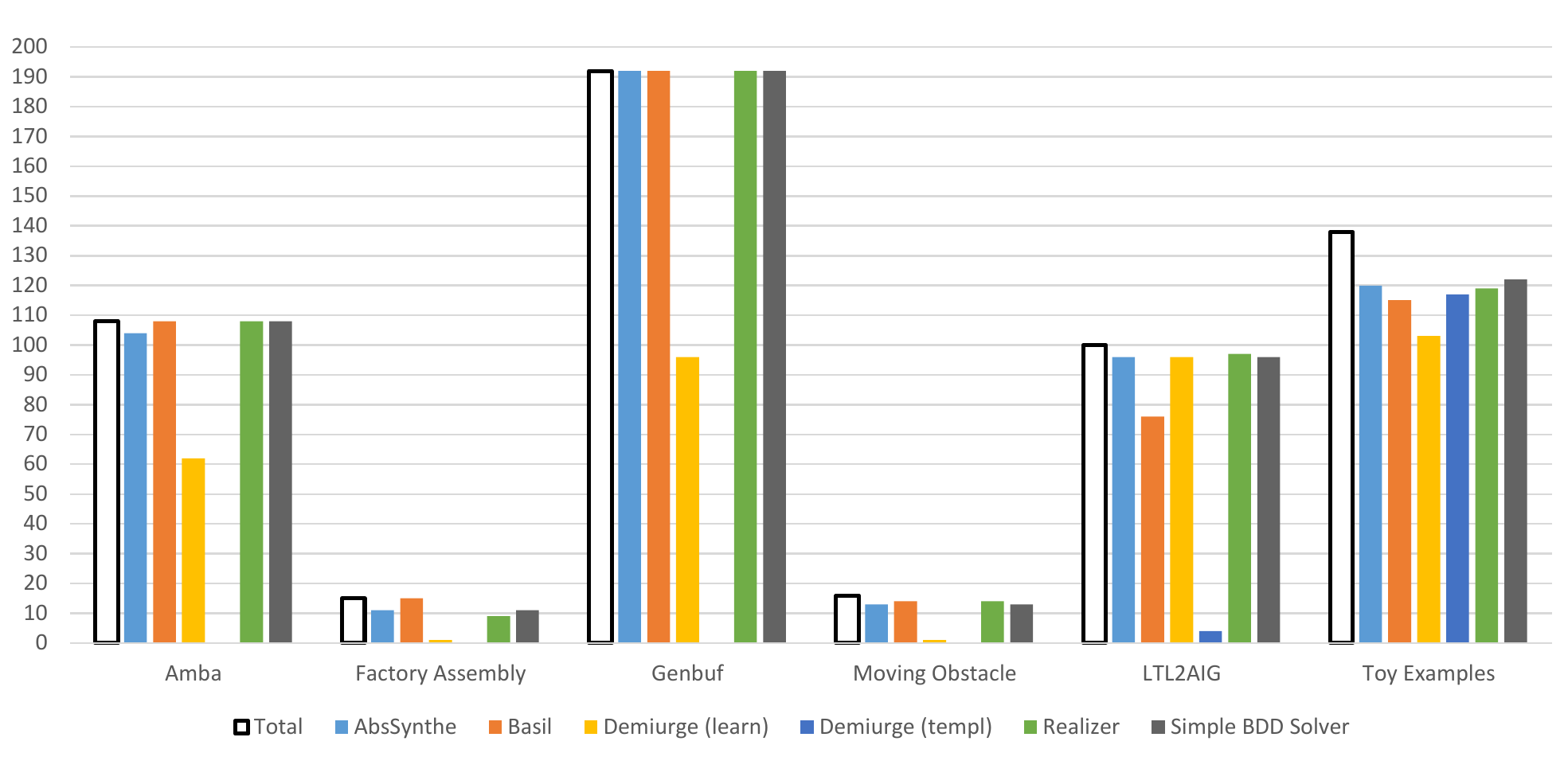}
\caption{Sequential Realizability Track, Solved Instances by Category}
\label{fig:res-overview-classes}
\end{figure*}

\begin{figure*}
\centering
\includegraphics[width=\linewidth]{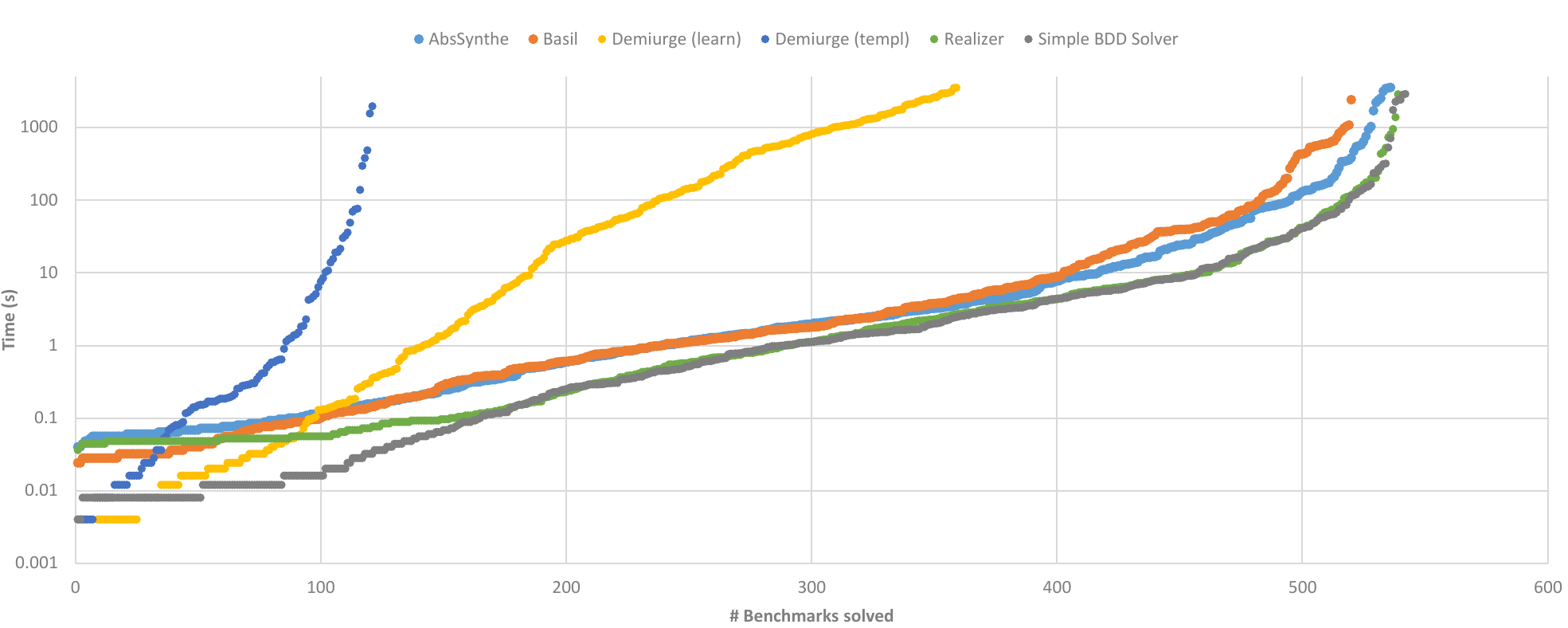}
\caption{Sequential Realizability Track, Cactus Plot}
\label{fig:rs-cactus}
\end{figure*}

Furthermore, Figure~\ref{fig:res-overview-classes} gives an overview of solved instances by
participant and benchmark classes (see Section~\ref{sec:benchmarks}), and
Figure~\ref{fig:rs-cactus} a cactus plot for the time needed by each tool to
solve the benchmarks.

\begin{table}
\caption{Results of the Sequential Realizability Track}
\label{tab:RealSeq}
\centering
\begin{tabular}{c|ccc}
\hline
Tool			& Solved  &  Unique & Relative \\ 
 \hline
\simpleBDD              	& {\bf 542} 	  & 0 & {\bf 262} 	\\
  \realizer 	 	 	& 539 	  & 1 & 229  \\
  \abssynthe 	 	   	& 536 	  & 0 & 144\\
\basil 		 	 	& 520 	  & 4 & 209  \\
\demiurge \learn	  	& 359 	  & 0 & 209  \\
\demiurge \templ	  	& 121 	  & {\bf 7} & 90  \\ \hline
\end{tabular}

\vspace{3pt}
The best result in each column is in bold.
\end{table}

\begin{table*}
\caption{Benchmark instances that were solved uniquely or not solved at all in at least one
  subtrack. ``Yes'' means solved by more than one tool, ``Yes$^*$'' means
  uniquely solved. ``-'' means that this benchmark instance was not tested.}
\label{tab:unsolved}
\begin{tabular}{r|c|c|c|c}
\hline
& Solved in &  Solved in & Solved in & Solved in \\
Benchmark & Seq. Realizability  & Par. Realizability &
Seq. Synthesis & Par. Synthesis\\
\hline
\texttt{amba8c7y} & Yes & Yes & No & No\\
\texttt{amba9c5y} & Yes & Yes & Yes & Yes$^*$\\
\texttt{amba10c5y} & Yes & Yes & Yes & Yes$^*$\\
\texttt{cnt30n} & No & No & - & -\\
\texttt{cnt30y} & No & No & No & No\\
\texttt{factory\_assembly\_5x5\_2\_10errors}  & Yes$^*$ & Yes$^*$ & No & No\\
\texttt{factory\_assembly\_5x5\_2\_11errors}  & Yes$^*$ & Yes$^*$ & No & No\\
\texttt{factory\_assembly\_7x5\_2\_10errors} & Yes$^*$ & Yes$^*$ & No & No\\ 
\texttt{factory\_assembly\_7x5\_2\_11errors} & Yes$^*$ & Yes$^*$ & No & No\\ 
\texttt{gb\_s2\_r2\_1\_UNREAL} & Yes$^*$ & No & - & -\\
\texttt{gb\_s2\_r2\_2\_REAL} & No & No & No & No\\ 
\texttt{gb\_s2\_r2\_3\_REAL} & No & No & No & No\\
\texttt{gb\_s2\_r2\_4\_REAL} & No & No & No & No\\
 \texttt{moving\_obstacle\_24x24\_7glitches} & Yes & Yes & No & No\\
\texttt{moving\_obstacle\_32x32\_11glitches} & Yes & Yes & No & No\\
\texttt{moving\_obstacle\_48x48\_19glitches} & Yes & Yes & No & No\\
\texttt{moving\_obstacle\_64x64\_27glitches} & Yes & Yes & No & No\\
\texttt{moving\_obstacle\_96x96\_43glitches} & Yes & Yes & No & No\\
\texttt{moving\_obstacle\_128x128\_59glitches} & No & No & - & -\\
\texttt{moving\_obstacle\_128x128\_60glitches} & No & No & - & -\\
\texttt{mult11} & Yes & Yes$^*$ & - & - \\
\texttt{mult12} & Yes$^*$ & No & No  & No\\
\texttt{mult13} & Yes$^*$ & No & -  & -\\ 
\texttt{mult14} & Yes$^*$ & No & -  & -\\ 
\texttt{mult15} & Yes$^*$ & No & -  & -\\ 
\texttt{mult16} & Yes$^*$ & No & No & No \\ 
\texttt{stay16y} & Yes & Yes$^*$ & Yes & Yes$^*$ \\
\texttt{stay18n} & Yes$^*$ & No & -  & -\\
\texttt{stay18y} & No & No & - & -\\
\texttt{stay20n} & Yes$^*$ & No & - & - \\
\texttt{stay20y} & No & No & - & -\\
\texttt{stay22n} & No & No & - & -\\ 
\texttt{stay22y} & No & No & - & -\\  
\texttt{stay24n} & No & No & - & -\\  
\texttt{stay24y} & No & No & - & -\\ 
\hline
\end{tabular}
\end{table*}

\myparagraph{Analysis} 
Table~\ref{tab:RealSeq} and Figure~\ref{fig:res-overview-classes} show that the
BDD-based tools \abssynthe, \basil, \realizer and \simpleBDD are very close to
each other when only comparing the number of instances solved. Furthermore, for
the Amba and Genbuf benchmarks, some tools solve all instances in the benchmark
set, i.e., we would
need more difficult instances to distinguish which tool is better for these
classes. 

Regarding the SAT- and QBF-based synthesis approaches, \demiurge \learn
solves about as many of the LTL2AIG benchmarks as the best BDD-based tools, and
almost as many of the Toy Examples. For AMBA and Genbuf Benchmarks, \demiurge
\learn solves only about half as many benchmarks, and for the Moving
Obstacle and Factory Assembly benchmarks can only solve one in each
case. Finally, \demiurge \templ can solve a very good number of the Toy
Examples, and even solves $7$ of them uniquely. However, it solves only very few
of the LTL2AIG benchmarks, and none of the others.

As can be seen in Figure~\ref{fig:rs-cactus}, most 
tools have a steep degradation with higher complexity of benchmarks, 
i.e., between $90\%$ and $95\%$ of the benchmarks that can be solved within 
the timeout of $5000$ seconds can actually be solved very quickly, i.e., in 
less than $600$ seconds. 

The uniquely solved benchmarks also show that there are significant differences 
between the algorithms of different tools. In particular, the template-based 
variant of \demiurge, while not very successful overall, can determine 
realizability for a relatively large number of Toy Examples that cannot be
solved by the other approaches.

Regarding the relative ranking, in Table~\ref{tab:RealSeq} we note that 
\demiurge \learn has the same score as \basil (and higher than \abssynthe), 
even though it can only solve $359$ problems, compared to $520$ for \basil 
and $536$ for \abssynthe. This is because this ranking rewards 
\demiurge for being one of the fastest tools on many of the small problem 
instances.

\mysubsubsection{Parallel Subtrack}
The parallel realizability subtrack had $4$ participants: parallel versions 
of \demiurge and \realizer\footnote{Due to a bug, the parallel version of 
\realizer   performed worse than the sequential version, as mentioned in   
Section~\ref{sec:participants}.}, and sequential versions of \abssynthe and 
\basil. The results are summarized in Table~\ref{tab:RealPar}. Again, no tool 
could solve all $569$ benchmarks. Table~\ref{tab:unsolved} shows $21$ 
benchmarks that were not solved by any of the tools within the timeout, and $6$
benchmarks that were solved uniquely by one tool. The successful tools are: 
\begin{itemize} \item \basil: $4$ versions of the factory assembly benchmarks 
(the same as before) \item \abssynthe: \texttt{mult11} and \texttt{stay16y} 
\end{itemize} We note that \realizer could not solve the problems that it 
uniquely solved in sequential execution mode. Furthermore, \demiurge \templ 
did not compete in the parallel subtrack, therefore its uniquely solved problems 
from the sequential subtrack are unsolved here.
 
\begin{figure*}
\centering
\includegraphics[width=\linewidth]{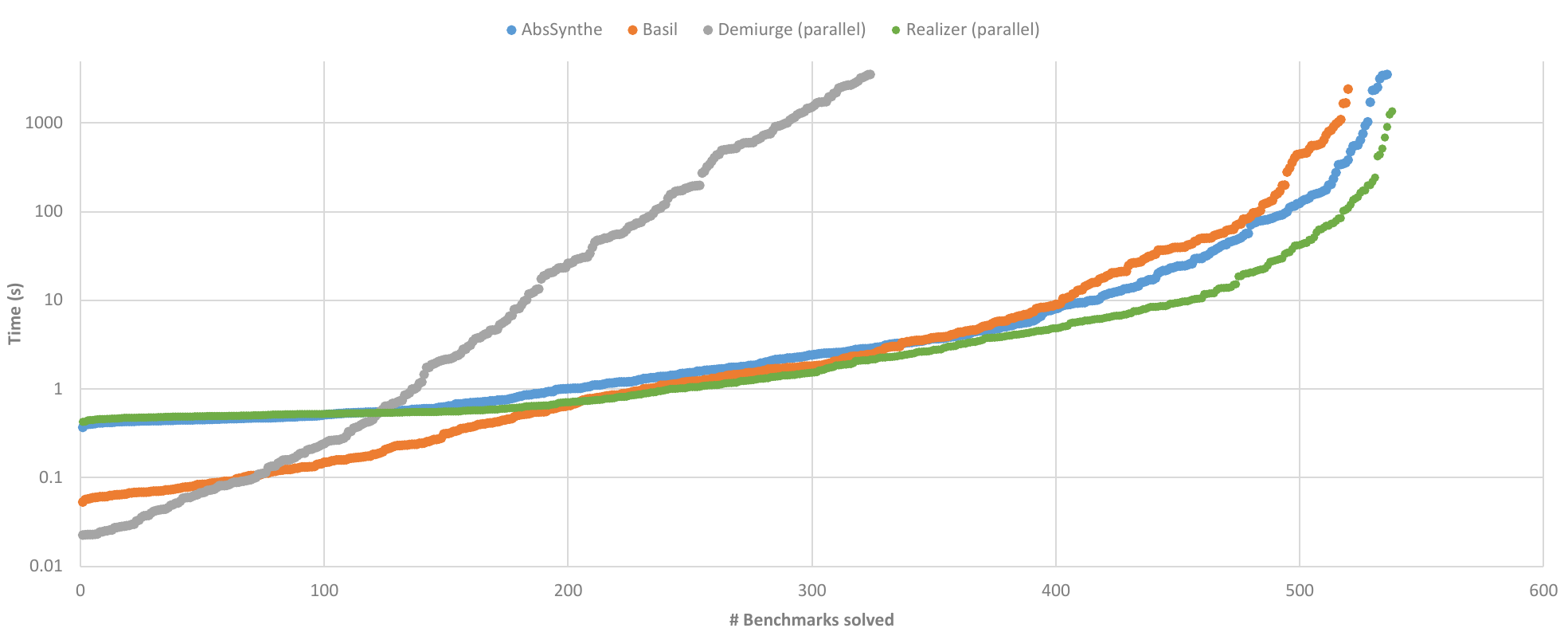}
\caption{Parallel Realizability Track, Cactus Plot}
\label{fig:rp-cactus}
\end{figure*}

A cactus plot for the number of benchmarks that can be solved by each tool 
within the timeout is given in Figure~\ref{fig:rp-cactus}. We do not give a
detailed analysis of the number of solved instances by category, since it is
very similar to the analysis in Figure~\ref{fig:res-overview-classes}.

\myparagraph{Analysis} 
Figure~\ref{fig:rp-cactus} shows the same steep degradation of
runtime with increasing complexity as in the sequential case.
Concerning the effectiveness of parallel versus sequential implementations,
this subtrack shows that currently none of the 
implementations benefits from using parallelism. To the 
contrary, the parallel implementations of \demiurge \learn and \realizer were not able to solve 
(in 5000s Wall time) the problems that their sequential implementations solved 
uniquely in the sequential subtrack (in 5000s CPU time), and \abssynthe and \basil are sequential implementations. 

Regarding the relative ranking based on Wall time, 
we note another weakness of the chosen ranking system:
this ranking heavily favors implementations in C++ that have a quick startup
time. \basil solves 150 problems in less than $0.36$s Wall time, which is the minimal time needed to solve any single problem for the Python-based implementations \abssynthe and \realizer. 
That is, the relative ranking scheme and our benchmark selection
favors tools that can solve easy benchmarks very quickly, and in particular 
the tools implemented in C++, as they make very efficient use of Wall time.

Summing up, we see that in the realizability tracks the BDD-based tools in
general outperform the SAT- and QBF-based 
approaches, except for a small subset of the benchmarks. Between the BDD-based
approaches, the differences we could detect are rather small --- the percentage
of benchmarks that can be solved by the BDD-based implementations ranges only
from 91\% to 95\%. None of the tools benefits from parallelism.

\begin{table}
\caption{Results of the Parallel Realizability Track}
\label{tab:RealPar}
\centering
\begin{tabular}{c|ccc}
\hline
Tool 			& Solved & Unique & Relative 	\\ \hline
\realizer		& {\bf 538}		 & 0 & 279 	 \\
\abssynthe & 536		 & 2 & 219 		\\
\basil 			& 520		 & {\bf 4} & {\bf 331} 	\\
\demiurge \dparallel		& 324		 & 0 & 133		\\\hline
\end{tabular}

\vspace{3pt}
The best result in each column is in bold.
\end{table}

\subsection{Synthesis Track}
In the synthesis track, tools were evaluated with respect to the relative and quality rankings (see Section~\ref{sec:rankings}), based on the size of solutions. Since these rankings are only defined on realizable specifications, we 
excluded all unrealizable specifications. Furthermore, we excluded most of the
problems that could not be solved by any tool in the realizability
track, since synthesis in general is harder than realizability checking. 
Out of the remaining $382$ benchmarks, we chose $157$ benchmarks with the 
goal to ensure a 
good coverage of different benchmark classes, and a good distribution over 
benchmarks of different difficulty. Only $3$ out of the $5$ tools that entered \syntcomp
competed in the synthesis track: \abssynthe, \basil and \demiurge.

\mysubsubsection{Sequential Subtrack}
The sequential synthesis subtrack had $4$ participants: \abssynthe and \basil
competed with one configuration each, and \demiurge with 
two different configurations.
Table~\ref{tab:SyntSeq} shows the number of solved
instances, and the accumulated points per tool in the relative and quality rankings. Note that a problem only counts as solved if the solution is successfully model checked. The number of model checking timeouts (MCTO) is also given in the table.

No tool could solve all $157$ benchmarks. No benchmark was solved 
uniquely 
by one tool, and $14$ benchmarks were solved by none of the tools 
(see Table~\ref{tab:unsolved}). \abssynthe solved the 
highest number of problems and earns the highest score in the quality ranking.
\demiurge \learn earns the highest score in our relative ranking, even though 
it solves less problems than \abssynthe.

\setlength{\tabcolsep}{.4em}
\begin{table}
\caption{Results of the Sequential Synthesis Track}
\label{tab:SyntSeq}
\centering
\begin{tabular}{c|ccccc}
\hline
Tool			& Solved  & Relative & Quality  & MCTO \\  \hline
\abssynthe 		     & {\bf 143}	& 329  & {\bf 265} & 6  \\
\demiurge \learn	    & 121	& {\bf 379} & 240 & 0   \\
\basil		     & 117	& 218  & 219 	& 5 \\
\demiurge \templ	   & 31 & 77 & 57 & 0          \\\hline
\end{tabular}

\vspace{3pt}
The best result in each column is in bold.
\end{table}

\begin{figure*}
\centering
\includegraphics[width=\linewidth]{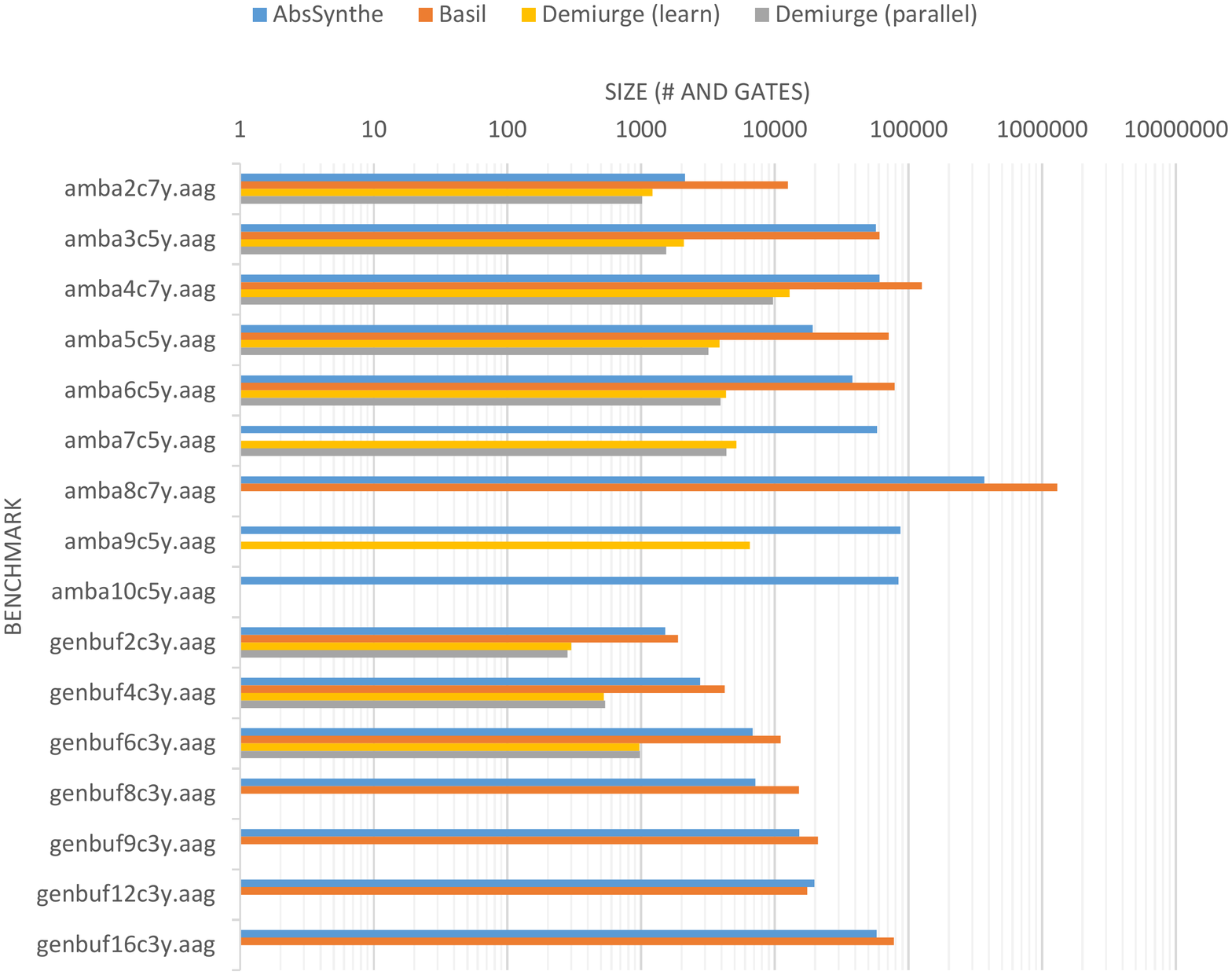}
\caption{Comparison of implementation sizes for a subset of the AMBA and GenBuf benchmarks}
\label{fig:size-ss2}
\end{figure*}

\begin{figure*}
\centering
\includegraphics[width=\linewidth]{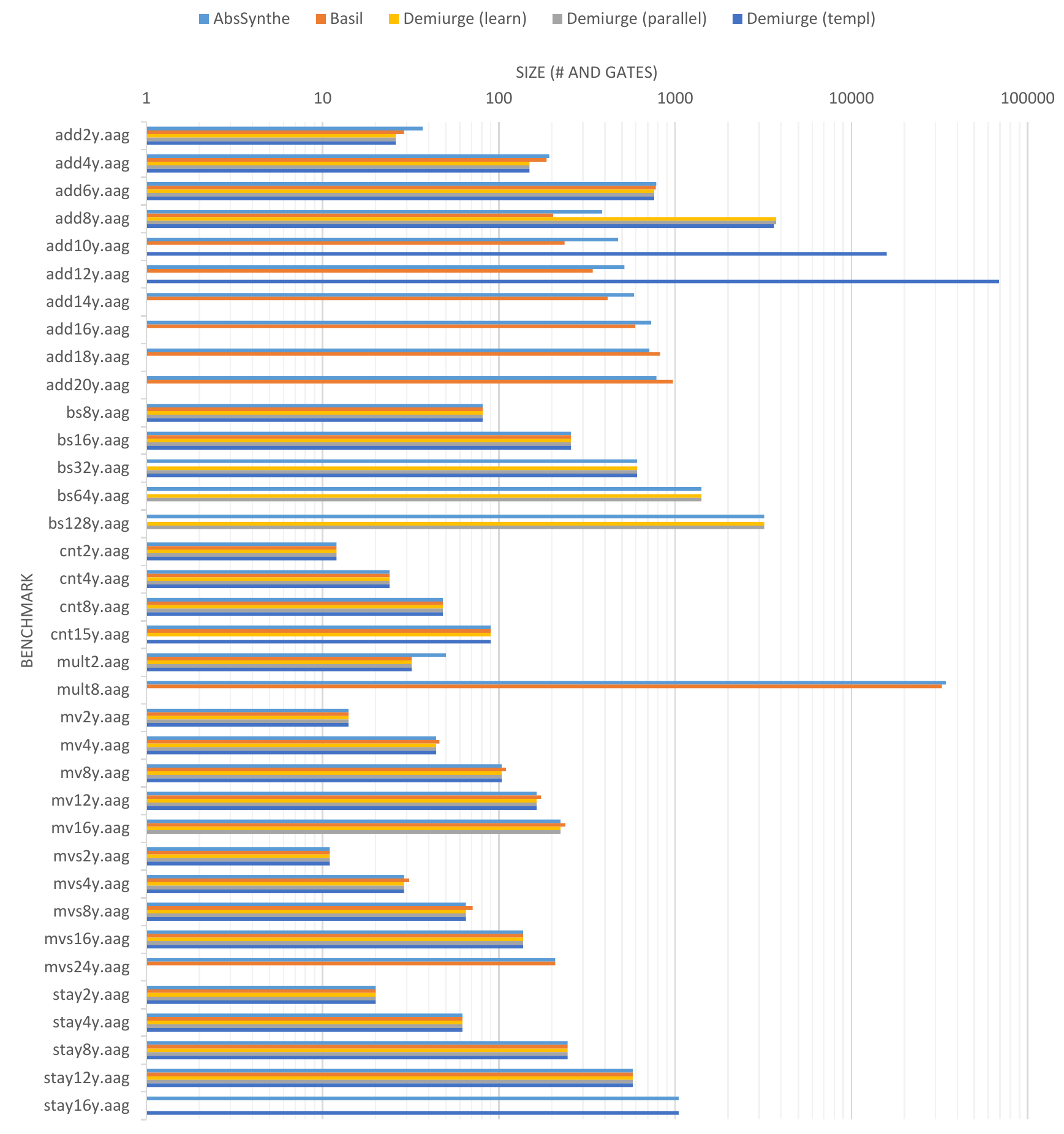}
\caption{Comparison of implementation sizes for a subset of the toy example benchmarks}
\label{fig:size-ss1}
\end{figure*}

Both \abssynthe and \basil produced a small number of 
solutions that could not be model checked within the separate $3600$s
timeout.
While counting these additional solutions would have 
changed the scores of these tools, it would not have 
changed the order of tools in any of the rankings.
Figures~\ref{fig:size-ss2} and~\ref{fig:size-ss1} give an 
overview of the size of produced implementations for a subset of the benchmarks,
showing significant differences on implementation sizes for some instances, in
particular from the AMBA and Genbuf classes.

\myparagraph{Analysis}
Regarding the relative and quality rankings, we note that \demiurge \learn profits from taking solution sizes into account. Figure~\ref{fig:size-ss2} shows that for 
those instances of the AMBA and GenBuf benchmarks that \demiurge \learn can 
solve, it provides implementations that are often by an order of magnitude 
smaller than those of the other tools. Figure~\ref{fig:size-ss1} shows a number of benchmarks where the implementation sizes are equal or very similar. Most of the time, the solutions of \demiurge \learn are smaller than those of \abssynthe and \basil, which is why it scores higher than \abssynthe and much higher than \basil in the relative ranking,
even though it solves less problems than \abssynthe, and about as many as
\basil. In the quality ranking, the relative difference between \demiurge \learn and \abssynthe is significantly smaller than in the number of solved instances ($9.5\%$ versus $15.5\%$ difference).

Furthermore, we note that the benchmark set contains 
relatively many problems that are easy to solve. For example, \abssynthe can 
solve $75$ of the $157$ problems in less than $0.5$s CPU time. 

Comparing the BDD-based tools, \abssynthe  
solves a number of problems that \basil cannot solve, and 
provides smaller solutions in many cases.

\mysubsubsection{Parallel Subtrack}
The parallel synthesis subtrack had $3$ participants: one configuration each of
\abssynthe, \basil, and \demiurge. \demiurge \dparallel was the only tool to use 
parallelism in the synthesis track.
The results are summarized in Table~\ref{tab:SyntPar}.
No tool could solve all $157$ benchmarks, $3$ benchmarks were solved uniquely 
by one tool, and $14$ benchmarks were solved by none of the tools.

\begin{table}
\caption{Results of the Parallel Synthesis Track}
\label{tab:SyntPar}
\centering
\begin{tabular}{c|ccccc}
\hline
Tool 							& Solved 	& Relative & Quality & MCTO	\\ \hline
\abssynthe 							& {\bf 143}			& 352 & {\bf 266} & 6\\
\demiurge \dparallel			& 119			& {\bf 393} & 237 & 0\\
\basil 							& 117		& 235 & 196	& 5\\ \hline
\end{tabular}

\vspace{3pt}
The best result in each column is in bold.
\end{table}

The benchmarks solved uniquely by one tool are:
\begin{itemize}
\item \abssynthe: \texttt{amba9c5y}, \texttt{amba10c5y}, and \texttt{stay16y}.
\end{itemize}

Like in the sequential subtrack, both \abssynthe and \basil produced a small 
number of solutions that could not be model checked. Implementation sizes for
\demiurge \dparallel are included in Figures~\ref{fig:size-ss2} and
\ref{fig:size-ss1}, showing that in some cases the implementations are even
smaller than those obtained from \demiurge \learn, in particular for the AMBA
benchmarks.

\myparagraph{Analysis}
Like in the sequential synthesis subtrack, \demiurge profits from providing small
solutions, even though it solves less problems than its competitors.
Furthermore, we note that \demiurge in this case profits from
parallelism to some extent. While it solved $2$ problems less than the
sequential \demiurge \learn, the solutions provided by \demiurge \dparallel were
in some cases even smaller than those provided by the sequential version.

\subsection{Observations on the State of the Art}

\mysubsubsection{BDD-based Synthesis} 
The standard BDD-based fixpoint algorithm for solving safety games is currently
the most efficient way for realizability checking based on monitor circuits. 
Implementations of the algorithm build on existing BDD packages, 
including operations for composition, abstraction, and dynamic reordering of 
BDDs. Based on these complex BDD operations, a competitive implementation can 
be fairly simple, as can be seen for example in \simpleBDD, which only consists
of about $300$ lines of code.
A few optimizations seem to be crucial, like automatic reordering, partitioned
transition relations,
and direct substitution. For other optimizations, like eager deallocation of
BDDs or simultaneous conjunction and abstraction, we have mixed results: the
tool authors that 
implemented them report increased efficiency, but we also have competitive
tools that do not implement them. 

A drawback of BDD-based synthesis becomes apparent when comparing
the size of solutions to those of \demiurge \learn: in many cases, the produced
implementations are much larger than necessary. 

As can be expected, a deeper analysis of the
runtime behavior of BDD-based tools shows that most of the time is spent
manipulating BDDs, in particular in the automatic reordering operations.
Therefore, it can be expected that the performance of BDD-based implementations  
heavily depends on the performance of the used BDD package.
Since all of the tools in \syntcomp 2014 use the same BDD package, the results of
the competition do not shed light on this issue, however.

\mysubsubsection{Template-based Synthesis} 
The template-based algorithm implemented in \demiurge \templ only solves a 
small subset of the benchmark set --- a closer analysis shows that it only 
performs well if a simple CNF representation of the winning region exists, 
which applies only to few \syntcomp benchmarks.  Hence, its performance on 
average is rather poor. However, this approach solves large instances of the 
\textsf{mult}, \textsf{cnt} and \textsf{stay} benchmarks much faster than the 
competition, or solves them uniquely.

\mysubsubsection{Learning-based Synthesis}
The learning-based algorithm implemented in \demiurge \learn solves far more
benchmarks than the template-based 
algorithm: $62\%$ of the benchmarks instead of $21\%$ in the sequential
realizability track. Still, the approach cannot really compete with the
BDD-based tools, which solve more than $90\%$. In the  
parallel realizability track, the situation is similar.

In the synthesis tracks, which are restricted to realizable problems and have rankings that take
into account the size of solutions, \demiurge \learn performed much better. 
Here, it solves $77\%$ of the benchmarks, compared to $78\%$ for \basil and
$95\%$ for \abssynthe (before model checking). Additionally, the learning-based algorithm produces
circuits that are sometimes several orders of magnitude smaller than those
produced by the BDD-based tools. This is also highlighted by
the fact that all solutions of \demiurge \learn are successfully model checked,
while both \abssynthe and \basil produce a number of solutions that can not be
verified within the timeout.

\mysubsubsection{Parallel Subtracks}
The submitted tools in general do not use parallelization very efficiently. The
parallel version of \realizer performs worse than the sequential version due to
a bug. For the parallel version of \demiurge, the result is double-edged: on the
one hand, the parallel version solves $2$ problems less than the sequential
version, on the other hand the solutions provided are often even smaller than
the ones produced by the sequential version.

For BDD-based tools, the lack of efficient parallel implementations 
correlates with the lack of efficient parallelized operations in BDD 
packages. While there have been recent efforts to parallelize BDD 
operations~\cite{DijkLP13,DijkP15}, this package does not support the important 
automatic reordering of BDDs, which makes it hard to integrate into a technique
that heavily relies on reordering.


%% file: conclusions.tex
\section{Conclusions and Future Plans}

\begin{sloppypar}
\syntcomp 2014 was a big success, making the first step towards establishing 
the competition as a regular event and its benchmark format as a standard 
language in the synthesis community. A number of synthesis tools have been 
developed specifically for the competition (\abssynthe, \basil, \realizer), 
while others are new versions or modifications of existing tools 
(\demiurge, \simpleBDD). Recently, the competition format has also been adopted by tool developers that have thus far not participated in \syntcomp~\cite{ChiangJ15}. Furthermore, the competition has sparked a lively 
discussion on the implementation of efficient synthesis techniques, in 
particular making tool developers aware of the range of optimizations used in 
BDD-based synthesis algorithms, and alternative SAT- and QBF-based approaches 
that are competitive at least on some classes of benchmarks.
\end{sloppypar}

At the time of this writing, \syntcomp 2015 has already been held~\cite{JacobsETAL16}. For the second iteration of the competition, we have expanded the benchmark set to more challenging
benchmarks, and to a wider range of different benchmark classes. Additionally,
following ideas of Sutcliffe and Suttner~\cite{SutcliffeS01} we have developed a classification scheme for benchmarks in terms of difficulty,
based on the results of \syntcomp 2014. Using this classification,
in \syntcomp 2015 we selected benchmarks to balance the weight
of benchmark instances from different classes and different difficulties.

Finally, recall that \syntcomp 2014 (and 2015) was restricted to the synthesis of finite-state systems from 
pure safety specifications in AIGER format. On the one hand, this resulted 
in a low entry-barrier for the competition and revived interest in the 
synthesis from pure safety specifications, as witnessed by several new tools 
and research papers related to the competition~\cite{BloemEKKL14,BloemKS14,BrenguierPRS14}. 
On the other hand, many of the existing synthesis tools did 
not participate because their strengths are in different kinds of synthesis 
tasks, for example in the synthesis from specifications in richer 
specification languages such as GR(1) or LTL. Thus, many interesting 
synthesis approaches are currently not covered by the competition. For 
\syntcomp 2016, we plan to extend the competition to a specification format 
that includes both GR(1) and LTL specifications~\cite{JacobsK16}.

{\small
\myparagraph{Acknowledgements}
We thank the anonymous reviewers for their detailed and insightful comments on drafts of this article. 
We thank Armin Biere for his advice on running a competition, and Ayrat Khalimov
for supplying the reference implementation Aisy for the competition.

\begin{sloppypar}
The organization of \syntcomp 2014 was supported by the Austrian Science Fund
(FWF) through projects RiSE (S11406-N23) and QUAINT (I774-N23), by the German
Research Foundation (DFG) as part of the Transregional Collaborative Research
Center ``Automatic Verification and Analysis of Complex Systems'' (SFB/TR 14
AVACS) and through project ``Automatic Synthesis of Distributed and
Parameterized Systems'' (JA 2357/2-1), as well as by the Institutional Strategy
of the University of Bremen, funded by the German Excellence Initiative.
The development of \abssynthe was supported by an F.R.S.-FNRS fellowship, and
the ERC inVEST (279499) project.
The development of \basil was supported by the Institutional Strategy
of the University of Bremen, funded by the German Excellence Initiative.
The development of \demiurge was supported by the FWF through projects RiSE
(S11406-N23, S11408-N23) and QUAINT (I774-N23).
The development of \realizer was supported by the DFG as part of SFB/TR 14
AVACS.
The development of \simpleBDD was supported by a gift from the Intel
Corporation, and NICTA is funded by the Australian Government through the
Department of Communications and the Australian Research Council through the ICT
Centre of Excellence Program.
\end{sloppypar}
}